\documentclass[conf]{new-aiaa}
\usepackage[table]{xcolor}
\usepackage{textcomp}
\usepackage{amsmath,amsfonts}
\usepackage{graphicx}
\usepackage{tikz}
\usepackage{url}

\usepackage{colortbl}
\usepackage{booktabs}
\usepackage{array}
\usepackage{tabularx}

\newcolumntype{L}[1]{>{\raggedright\let\newline\\\arraybackslash\hspace{0pt}}m{#1}}

\DeclareMathOperator*{\argmin}{argmin}

\newcommand{\mc}[1]{\ensuremath{\mathcal{#1}}}   

\newcommand{\nth}[1]{\ensuremath{#1^\text{th}}}

\makeatletter
\DeclareRobustCommand\onedot{\futurelet\@let@token\@onedot}
\def\@onedot{\ifx\@let@token.\else.\null\fi\xspace}

\usepackage[T1]{fontenc}  
\usepackage[b]{esvect}    

\makeatletter
\newlength\xvec@height%
\newlength\xvec@depth%
\newlength\xvec@width%
\newcommand{\xvec}[2][]{%
  \ifmmode%
    \settoheight{\xvec@height}{$#2$}%
    \settodepth{\xvec@depth}{$#2$}%
    \settowidth{\xvec@width}{$#2$}%
  \else%
    \settoheight{\xvec@height}{#2}%
    \settodepth{\xvec@depth}{#2}%
    \settowidth{\xvec@width}{#2}%
  \fi%
  \def\xvec@arg{#1}%
  \def\xvec@dd{:}%
  \def\xvec@d{.}%
  \raisebox{.2ex}{\raisebox{\xvec@height}{\rlap{%
    \kern.05em
    \begin{tikzpicture}[scale=1]
    \pgfsetroundcap
    \draw (.05em,0)--(\xvec@width-.05em,0);
    \draw (\xvec@width-.05em,0)--(\xvec@width-.15em, .075em);
    \draw (\xvec@width-.05em,0)--(\xvec@width-.15em,-.075em);
    \ifx\xvec@arg\xvec@d%
      \fill(\xvec@width*.45,.5ex) circle (.5pt);%
    \else\ifx\xvec@arg\xvec@dd%
      \fill(\xvec@width*.30,.5ex) circle (.5pt);%
      \fill(\xvec@width*.65,.5ex) circle (.5pt);%
    \fi\fi%
    \end{tikzpicture}%
  }}}%
  #2%
}
\makeatother

\makeatletter
\renewcommand*\env@matrix[1][\arraystretch]{%
  \edef\arraystretch{#1}%
  \hskip -\arraycolsep
  \let\@ifnextchar\new@ifnextchar
  \array{*\c@MaxMatrixCols c}}
\makeatother

\newcommand{\FrameSymbol}{\mc{F}}                       
\newcommand{\frm}[1]{\mc{#1}}                           
\newcommand{\Frm}[1]{{{\FrameSymbol}^{\frm{#1}}}}       
\newcommand{\mat}[1]{{\mathbf{#1}}}                     
\newcommand{\point}[1]{{#1}}

\definecolor{commentcolor}{gray}{0.5}
\usepackage{algorithm}
\usepackage{algpseudocode}

\algnewcommand{\LineComment}[1]{\State \textcolor{commentcolor}{\(\triangleright\) #1}}
\algnewcommand{\To}{\textbf{to}}
\algnewcommand{\Break}{\textbf{break}}
\algnewcommand{\Continue}{\textbf{continue}}
\algnewcommand{\IIf}[1]{\State\algorithmicif\ #1\ \algorithmicthen}
\algnewcommand{\EndIIf}{\unskip}
\algnewcommand{\var}[1]{\textit{#1}}
\algnewcommand{\func}[1]{\textsc{#1}}

\newcommand{\ZI}{\hat{e}_d}

\newcommand{\ZR}[1]{\hat{u}_{#1}}

\newcommand{\Force}[1]{\mat{F}^\frm{B}_{#1}}
\newcommand{\Moment}[1]{\mat{M}^\frm{B}_{#1}}
\newcommand{\Thrust}[1]{\mat{T}_{#1}}
\newcommand{\Torque}[1]{\mat{\tau}_{#1}}

\newcommand{\RBI}{\mat{R}^{\frm{B}}_{\frm{I}}}

\usepackage[utf8]{inputenc}
\usepackage{multirow}
\usepackage{graphicx}
\usepackage{amsmath}
\usepackage[version=4]{mhchem}
\usepackage{siunitx}
\usepackage{longtable,tabularx}
\usepackage{epstopdf}
\usepackage{setspace}
\usepackage{mathtools}
\usepackage{soul}
\usepackage{grffile}
\usepackage{threeparttable}
\usepackage{gensymb}
\usepackage{subfigure}
\usepackage{todonotes}
\usepackage{diagbox}
\usepackage[absolute,overlay]{textpos}

\title{A Unified MPC Strategy for a Tilt-rotor VTOL UAV Towards Seamless Mode Transitioning}

\author{Qizhao Chen\footnote{Master Student, Department of Mechanical Engineering, qizhaoc@andrew.cmu.edu}, Ziqi Hu\footnote{Co-first author,Master Student, Department of Mechanical Engineering, ziqih@andrew.cmu.edu}}
\affil{Carnegie Mellon University, Pittsburgh, PA, 15213}
\author{Junyi Geng\footnote{Assistant Professor, Department of Aerospace Engineering, jgeng@psu.edu}}
\affil{Pennsylvania State University, University Park, PA 16802}
\author{Dongwei Bai\footnote{Master Student,Department of Mechanical Engineering, dongweib@andrew.cmu.edu},Mohammad Mousaei\footnote{Ph.D. Student, The Robotics Institute, mmousaei@andrew.cmu.edu},Sebastian Scherer \footnote{Associate Research Professor, The Robotics Institute, basti@andrew.cmu.edu}}
\affil{Carnegie Mellon University, Pittsburgh, PA, 15213}

\begin{document}
\begin{textblock*}{15cm}(4.75cm,1cm) 
   {\Large\textcolor{red}{AIAA SciTech Forum, January 8-12 2024}\\
   \textcolor{purple}{\url{https://arc.aiaa.org/doi/abs/10.2514/6.2024-2878}}}
\end{textblock*}

\maketitle

\begin{abstract}
Capabilities of long-range flight and vertical take-off and landing (VTOL) are essential for Urban Air Mobility (UAM). Tiltrotor VTOLs have the advantage of balancing control simplicity and system complexity due to their redundant control authority. Prior work on controlling these aircraft either requires separate controllers and switching modes for different vehicle configurations or performs the control allocation on separate actuator sets, which cannot fully use the potential of the redundancy of tiltrotor. This paper introduces a unified MPC-based control strategy for a customized tiltrotor VTOL Unmanned Aerial Vehicle (UAV), which does not require mode-switching and can perform the control allocation in a consistent way. The incorporation of four independently controllable rotors in VTOL design offers an extra level of redundancy, allowing the VTOL to accommodate actuator failures. The result shows that our approach outperforms PID controllers while maintaining unified control. It allows the VTOL to perform smooth acceleration/deceleration, and precise coordinated turns. In addition, the independently controlled tilts enable the vehicle to handle actuator failures, ensuring that the aircraft remains operational even in the event of a servo or motor malfunction. 


\end{abstract}

\section{Introduction} \label{sec:intro}
In recent years, Unmanned Aerial Vehicles (UAVs) have experienced extensive utilization across various fields and industries because of their flexibility and versatility, such as photography and 3D mapping~\cite{zhao2021super}, delivering package\cite{geng2020cooperative, geng2021estimation, geng2022load}, physical interaction, and aerial manipulation~\cite{he2023image, guo2023aerial}. UAVs with fixed rotors installed vertically have the ability to vertical take-off and landing (VTOL) and can hover, which is essential in many applications~\cite{Mousaei:2022:icra-workshop:vtol}. However, they are not energy efficient in the forward flight because they need to provide thrust which is used to counter gravity. Vehicles with fixed wings (e.g., airplanes) are way more energy efficient during cruise flight leading to much longer ranges than multirotors. The issue for them is that runways are usually required for take-off and landings, thus lacking VTOL capabilities. Hybrid VTOL UAVs, classified as the third category, solve these issues by integrating VTOL capabilities while being capable of efficient long-range flight. This makes them the leading contenders for Urban Air Mobility (UAM)\cite{mousaei2022design}.

\begin{figure}
    \centering
    \includegraphics[width=0.6\textwidth]{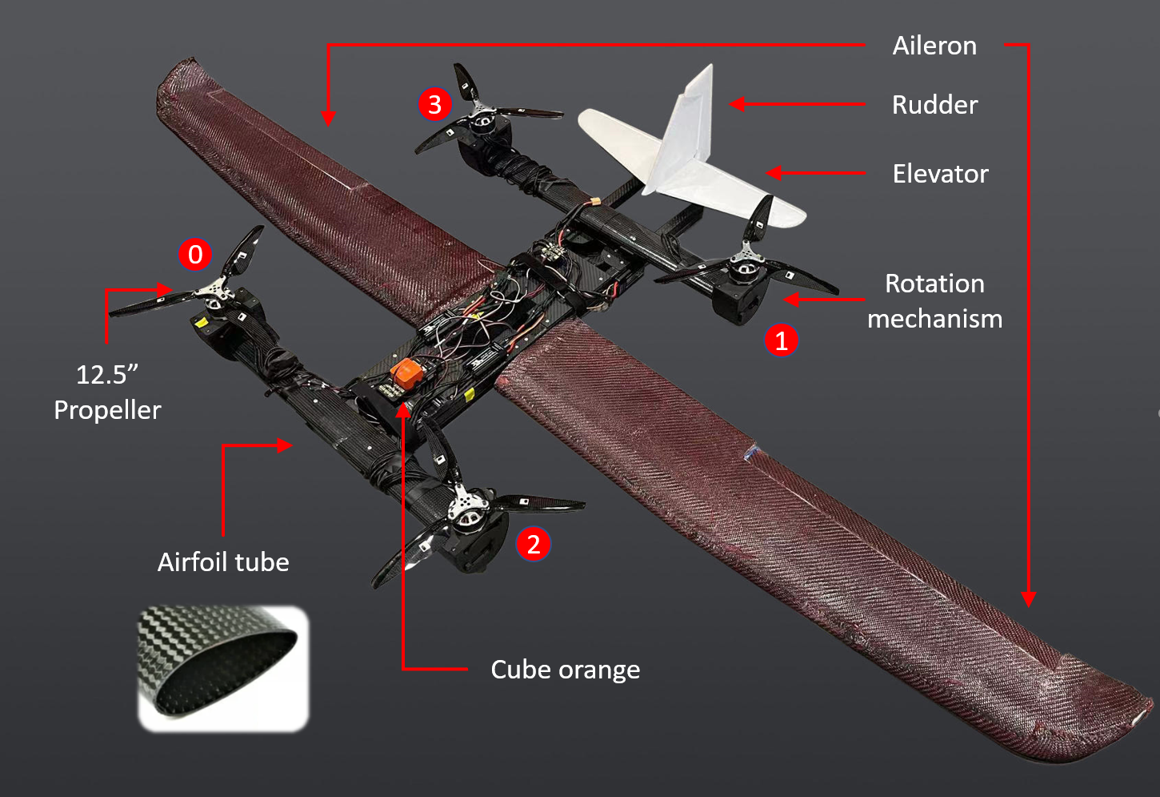}
    \caption{
The design of our VTOL tiltrotor aircraft includes four individual rotors, a tilting mechanism for the rotors, two ailerons, an elevator, and a rudder~\cite{mousaei2022design}. Notably, the quadrotor arms employ airfoil cross-sections.}
    \label{fig:platform}
\end{figure}

Researchers typically explore and investigate three different designs of VTOLs: standard fixed-rotor, tailsitters, and tiltrotors VTOLs, but striking a balance between control simplicity and configuration complexity can be challenging. Among these types, standard fixed-rotor VTOLs are generally considered the easiest to control. Nevertheless, incorporating separate propulsion systems for hover and cruise flight introduces an additional weight burden. Although the tailsitter VTOLs require the least number of actuators, it is difficult to control them due to the requirement of the rotating fuselage, especially in windy conditions. Tiltrotors offer improved control during hover compared to tailsitters due to their redundant control authority. However, the inclusion of a tilting mechanism adds complexity to the system. This paper focuses on a custom-designed tiltrotor VTOL UAV~\cite{mousaei2022design}, which reduces the system complexity, see Fig.\ref{fig:platform}. 

There are several challenges to controlling a hybrid VTOL UAVs, especially for the tiltrotors: (1) The control mechanisms are different for each VTOL's configuration or flight phase (multirotor, fixed-wing), e.g. for a climb up, the multirotor needs to pitch down while the fixed-wing needs to pitch up. (2) The transition phase is difficult to control due to the drastic structural changes and the complex aerodynamics induced by the rotor wing interference. (3) The over-actuated design, such as the tiltrotors raises the question of how to perform the control allocation to better utilize the configuration redundancy so that the transition performance and the overall efficiency can be improved.


A growing number of research appear to investigate the control of hybrid VTOL UAVs ranging from different vehicle designs~\cite{li2020transition, bauersfeld2021mpc}, different flight phases~\cite{xi2021design, reinhardt2021control}, to different control methods~\cite{hernandez2015transition, singh2021quadplus}. A prevalent approach for controlling VTOL UAVs involves utilizing three distinct controllers for each flight phase, including the multirotor phase, transition phase, and fixed-wing phase. A global task planner is typically employed to facilitate the transition among these controllers. One simple and intuitive solution is to blend the control from two flight configurations using a pre-planned feed-forward controller, usually PID for transition. This is also the way Pixhawk autopilot uses. However, such kind of open-loop controller is not robust and requires fine-tuning the gained airspeed and vehicle pose to achieve a successful transition. In order to improve the transition performance, researchers explored several control approaches besides PID, such as gain scheduling~\cite{hernandez2015transition}, adaptive control~\cite{marshall2022model}, sliding mode~\cite{yu2021immersion} and backstepping control~\cite{kong2018mathematical}. However, most of them treat complex aerodynamics as a bounded disturbance and rely on system robustness to handle it leading to conservative performance or even insuppressible behavior when the disturbance is large.

Recently, model predictive control has gained considerable popularity in the aerospace community, primarily due to its lower dependence on high-precision system models, making it a more flexible and adaptive control strategy.~\cite{prach2018mpc, reinhardt2021control}. Given the predictive dynamic model, it provides the optimal input for the system based on the state and input constraints while minimizing some pre-defined costs. Such kind of method relies on the optimization capability to compensate for the system uncertainty caused by the model mismatch, disturbance, etc. Some researchers utilize this approach on VTOL UAVs, such as tailsitters, however, less effort has been put into the tilt-rotor VTOL UAVs due to challenges stemming from its structural complexity and the redundant control authority. The paper by Bauersfeld et al. \cite{bauersfeld2021mpc} introduces an integrated controller capable of effectively controlling tiltrotor aircraft across the entire flight envelope, eliminating the need for switching between different flight phases. However, this work only utilizes the MPC for the velocity controller and still treats the aerodynamic actuators, tilt angles, and propellers separately in control allocation. An MPC-based attitude controller for a tilt-rotor VTOL UAV is designed in \cite{xi2021design} for the transition mode to predict the rotor/wing disturbance and complete the control by compensation. However, it focuses on a linearized model and still treats different flight phases separately which requires mode switching. 

This paper presents a novel unified Model Predictive Control (MPC)-based approach for a customized tiltrotor VTOL UAV, eliminating the need for mode switching and leveraging configuration redundancy. Initially, we introduce the design and model of our unique tiltrotor VTOL, which combines the distinguishing characteristics of a fixed-wing aircraft with a variable-pitch quadrotor UAV. Unlike previous designs such as those in \cite{bauersfeld2021mpc, ducard2014modeling}, our aircraft design features detached quadrotor arms and incorporates a controllable propeller direction and tilting mechanism at the ends of the arms. The airfoil cross-sections of the arms provide additional lift during forward motion. We also develop a nonlinear dynamics model for this tiltrotor VTOL. Next, we design an MPC-based controller for the velocity loop, offering a unified control approach. Unlike \cite{bauersfeld2021mpc}, our control allocation strategy optimizes the entire actuator set simultaneously rather than in separate groups. This integrated control framework enhances the efficiency and effectiveness of the control system. In addition, The integration of four independently controllable rotors in VTOL systems provides a range of substantial advantages. This configuration enhances maneuverability by enabling precise adjustments for complex flight patterns. Furthermore, the system ensures improved stability during both takeoff and cruise phases through the independent control of each rotor. The inclusion of four rotors also offers a crucial redundancy factor, enhancing overall safety by compensating for potential actuator failures. 

Our main contributions include:
\begin{enumerate}
    \item Propose a unified MPC-based control approach for a tiltrotor VTOL that can operate in all flight phases. The proposed method utilizes MPC for the velocity loop.
    \item Design a control allocation method so that the whole actuator set can be optimized at the same time. 
    \item Validate the system performance on a wide range of flight scenarios as well as the scenarios with different actuator failures.
    \item Provide the source code. The result is implemented on a customized multi-purpose simulator~\cite{Keipour:2022:unpub:simulator} which can be used by researchers for rapid prototyping other vehicle designs or controller development. 
\end{enumerate}

\section{Model} \label{sec:problem}
This section introduces the custom-designed tiltrotor VTOL UAV and the vehicle dynamics model. More details can be found in \cite{mousaei2022design}.

\subsection{Design Overview}

\begin{figure}
    \centering
    \includegraphics[width=0.6\linewidth]{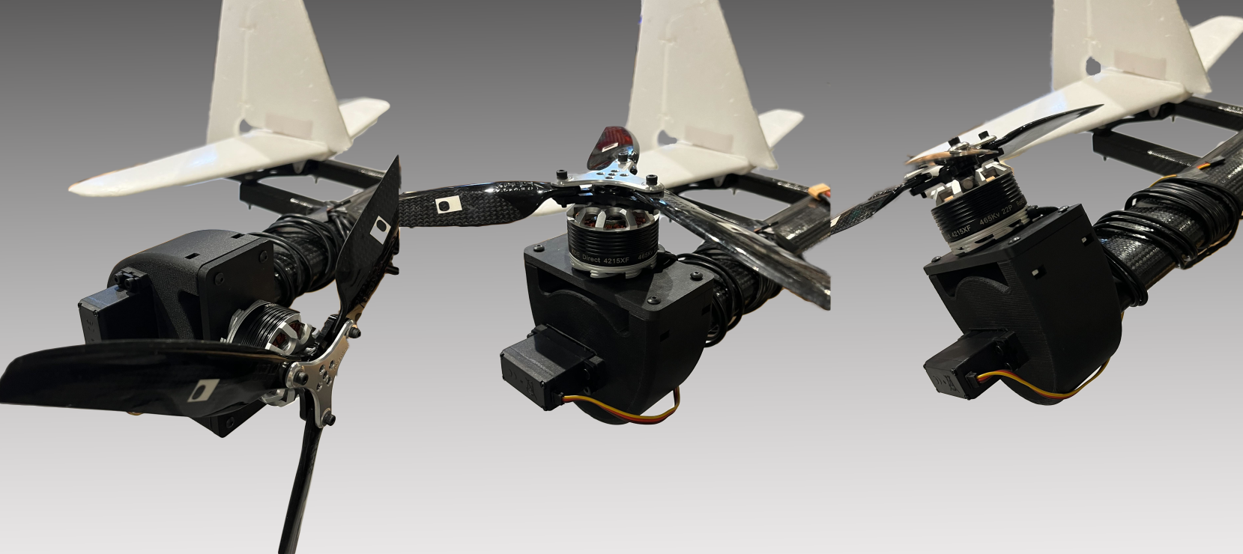}
    \caption{Rotor tilt in three different positions~\cite{mousaei2022design}}
    \label{fig:servo-tilt}
\end{figure}

The aircraft features vertically integrated quadrotor arms, which take the form of two airfoil tubes integrated into the fuselage of the aircraft. Thus our design completely separates the quadrotor arm from the wing side. We directly attach four motors to the ends of the arms. These motors are capable of driving a 12.5" propeller and delivering a thrust of approximately 27.36 N at 100\% throttle. A servo is utilized to power the rotation mechanism and connect the motor and propeller to the fuselage, which enables the necessary rotational movement for controlling the propeller's tilt, see Fig.~\ref{fig:servo-tilt}.

The fuselage is equipped with a set of wings that have a wingspan of approximately 2 meters. Additionally, the extended structure from the fuselage accommodates the installation of the elevator and rudder, serving as crucial control surfaces for pitch and yaw control, respectively. Fig.~\ref{fig:platform} shows the aircraft with the extended structure

Table \ref{tab:my_label} shows the vehicle geometry and inertial parameters.

\begin{table}[!htb]
    \centering
    \caption{Nomenclature}
    \label{tab:my_label}
    \begin{tabular}{ccc}
        \hline\hline
        \textbf{Name} & \textbf{Parameter} & \textbf{Value}    \\ \hline
        tilt angle &  $\mathcal{X}$&  $-7^\circ \leq \mathcal{X} \leq 95^\circ$    \\
        aileron & $\delta_{a}$ &  $\delta_{a} \in [-1, 1]$    \\
        elevator & $\delta_e$ &  $\delta_e \in [-1, 1]$    \\
        rudder & $\delta_r$ &  $\delta_r \in [-1, 1]$   \\
        airspeed & $V_a$ &  $0 \leq V_a \leq 30 m/s$\\
        mass & $m$ &  $7.427kg$    \\
        inertia & I &  $diag([10.685, 5.7465, 4.6678])kg. m^{2}$\\
        Inertial Frame & $\mathcal{I}$ &  $\{\point{O}_{\frm{I}}, e_n, e_e, e_d\}$    \\
        Body Fixed Frame & $\mathcal{B}$ &  $\{\point{O}_{\frm{W}}, e_x, e_y, e_z\}$    \\\hline
    \end{tabular}
\end{table}

\subsection{Definitions}
In this work, the inertial frame is established as: 
$\mathcal{F}^{I}=\{\point{O}_{\frm{I}}, \hat{e}_n, \hat{e}_e, \hat{e}_d\}$, 
where $\point{O}_{\frm{I}}$ is the origin,  $\hat{e}_n$, $\hat{e}_e$, and $\hat{e}_d$ represent unit vectors which are consistent with NED (North, East, Downwards) respectively. We define the body frame as:
$\mathcal{F}^{B}=\{\point{O}_{\frm{B}}, \hat{e}_x, \hat{e}_y, \hat{e}_z\}$, 
where $\point{O}_{\frm{B}}$ is the origin, and $\hat{e}_x$, $\hat{e}_y$, and $\hat{e}_z$ represent unit vectors pointing to the front, right, bottom directions respectively. The body frame is defined as:
$\mathcal{F}^{W}=\{\point{O}_{\frm{W}}, \hat{x}_W, \hat{y}_W, \hat{z}_W\}$, 
where $\point{O}_{\frm{W}}$ is arbitrary. The positive direction of the $\hat{x}_w$ axis corresponds to the direction of motion, $\hat{z}_w$ is perpendicular to plane $\hat{x}_w$ $\hat{y}_w$ and positive direction pointing to the bottom of the UAV, and $\hat{y}_w$ is perpendicular to 
the $\hat{x}_w\hat{z}_w$ plane following the right-hand rule. The angle of attack $\alpha$ is the angle between the body velocity vector and body reference plane  $\hat{x}_w\hat{y}_w$.
The sideslip angle is defined as the angle formed between the instantaneous direction of the velocity vector and its projection on the $\hat{e}_x \hat{e}_y$ plane. $R_{\mathcal{B}}^{\mathcal{I}}$ is the rotations matrix from body frame $\Frm{B}$ to inertial $\Frm{I}$.

The system comprises 12 actuators, each characterized by specific parameters. The deviation angles of the left and right ailerons from their neutral positions are denoted as $\delta_{a_1}$ and $\delta_{a_2}$ respectively. The deviation angle of elevation from the neutral position is represented by $\delta_e$, while the rudder's deviation angle is denoted as $\delta_r$. The rotational velocity of the  \nth{i}  rotor is denoted as $\omega_i$, and the corresponding rotation angle for each rotor is described by $\chi_i$, which represents the angle of the  \nth{i} rotor measured from $\hat{e}_x$ axis, with a value of 0 degrees when the rotor points upward and 90 degrees when it points to the front of the vehicle. 

\subsection{Thrust Forces and Moments}
\subsubsection{Thrust Forces}

The thrust $\Thrust{i}$ and torque $\boldsymbol{\Torque{i}}$ generated by the \nth{i} rotor can be approximated as:
\begin{align}
    \|\Thrust{i}\| = T_i = c_F \omega_i^2, \quad
    \|\boldsymbol{\Torque{i}}\| = {\tau_i} = (-1)^{d_i} c_K \omega_i^2
\end{align}
where $c_F$ and $c_K$ are the thrust coefficient and torque coefficient of each rotor and $d_i$ is the rotation in either a clockwise or counter-clockwise direction.\
The forces acting on the vehicle resulting from the propellers can be expressed as:
    \begin{align}
        \mathbf{F}_r^{\mathcal{B}} = \sum_{i = 1}^4 \mathbf{T}_i^{\mathcal{B}} + \mathbf{D}_i^{\mathcal{B}}, \quad
    \Thrust{i}^\frm{B} =
    T_i\ZR{i}^\frm{B} =
    c_F\omega_i^2 \begin{bmatrix}\sin {\chi_i}\\0\\-\cos{\chi_i}\end{bmatrix}
\end{align}
    where $\ZR{i}^\frm{B}$ is the direction of the force in body fixed frame and the tilt rotation matrix is
    \begin{equation}
        \mathbf{R}_{\mathcal{X}_i} = \begin{bmatrix}\cos \mathcal{X}_i & 0 &  -\sin \mathcal{X}_i \\0 & 1 & 0 \\ \sin \mathcal{X}_i & 0 & \cos \mathcal{X}_i\end{bmatrix}
    \end{equation}
    with the thrust constant being relatively small. The induced drag $\mathbf{D}_i^{\mathcal{B}}$ can be neglected, and
    \begin{equation*}
        \mathbf{F}_r^{\mathcal{B}} = \sum_{i = 1}^4 c_F\omega_i^2\begin{bmatrix}\sin \mathcal{X}_i\\0\\-\cos\mathcal{X}_i\end{bmatrix}
    \end{equation*}
    
    \subsubsection{Thrust Moments}
   Thrust moments arise from setting the rotor thrust with an offset from the center of gravity. These moments can be expressed in a simplified manner as:
    \begin{equation}
        \mathbf{M}_r^{\mathcal{B}} = \sum_{i=1}^4 (\mathbf{r}_i^{\mathcal{B}} \times \mathbf{T}_i^{\mathcal{B}})
    \end{equation}
    where $\mathbf{r}_i^{\mathcal{B}}$ is the distance between center of gravity to the \nth{i} rotor's position.
    
    \subsection{Aerodynamics Forces and Moments}
    The aerodynamic characteristics of the VTOL were obtained from an open-source program based on MATLAB called Tornado. The simulation requires geometry setup, flight condition setup, and lattice generation. The geometry setup requires details of the wings including wings' airfoil, chord length, span, etc. The flight condition setup requires $\alpha$ (angle of attack), $\beta$ (angle of sideslip), and angular velocity in roll, pitch, and yaw. Visualization of the wings setup is shown in the figure below. All aerodynamic characteristics have been exported for further calculation.
    

    \begin{figure}[hbt!]
        \centering
        \includegraphics[width=230px]{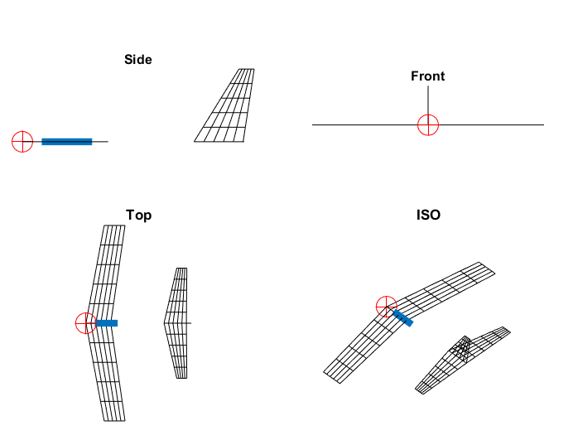}
        \caption{Geometry setup in different views}
        \label{fig:my_label}
    \end{figure}

     \subsubsection{Aerodynamic Force}
      The aerodynamic force $\mathbf{F}_a$ can be expressed in a simplified form as:
    \begin{align}
        \mathbf{F}_a^{\mathcal{B}} = \begin{bmatrix}X^{\mathcal{B}}\\Y^{\mathcal{B}}\\Z^{\mathcal{B}}\end{bmatrix}= \mathbf{R}_{\mathcal{W}}^\mathcal{B}\begin{bmatrix}X^\frm{W} \\Y^\frm{W} \\ Z^\frm{W}\end{bmatrix}
    \end{align}
    where X, Y, and Z represent drag, lateral, and lift force respectively. Notice the drag, lateral, and lift forces here exported from the simulation are in the wing's frame and being transformed into the body frame.
    \begin{align}
        X^{\mathcal{W}} &= \Bar{q} S C_X(\alpha, \beta) ~~~\mbox{(Drag Force)}\\
        Y^{\mathcal{W}} &= \Bar{q} S C_Y(\beta) ~~~\mbox{(Lateral Force)}\\
        Z^{\mathcal{W}} &= \Bar{q} S C_Z(\alpha) ~~~\mbox{(Lift Force)} 
    \end{align}
    with $\Bar{q} = \frac{\rho V_a^2}{2}$ as the dynamic pressure. $S$ is the wing surface area which is 0.44m$^2$ for our VTOL. $C_X$,$C_Y$, and $C_Z$ are the drag, lateral, and lift coefficient. When the lateral force is ignored, the lift and drag coefficient for the vehicle can be represented as:
    \begin{align}
        C_X(\alpha, \beta) &= C_D(\alpha) + C_{X\beta2}\beta^2 \approx C_D(\alpha) = C_{D0} + C_{D\alpha} \alpha^2 \\
        C_Z(\alpha) &= C_L(\alpha) = C_{Z0} + C_{Z\alpha} \alpha^2
    \end{align}
    where $C_{D0}$ = 0.35, $C_{D\alpha}$ = 0.11, $C_{Z0}$ = 0.03, $C_{Z\alpha}$ = 0.2, and -10$^\circ$$\le$ $\alpha$ $\le$10$^\circ$\\ 
    
    At low airspeed, VTOL tends to operate in a multirotor mode. The rotor thrust force and moment will dominate instead of the aerodynamic force and moment of the wing. We therefore use a constant $e$ to determine the effectiveness of aerodynamics. The equation for the aerodynamic force becomes:
    
    \begin{align}
        \mathbf{F}_a^{\mathcal{B}} = {e}\mathbf{F}_a^{\mathcal{B}}
    \end{align}
    where $e$ is determined by the airspeed. Their relationship is shown in the figure below.
    
    \begin{figure}[hbt!]
        \centering
        \includegraphics[width=230px]{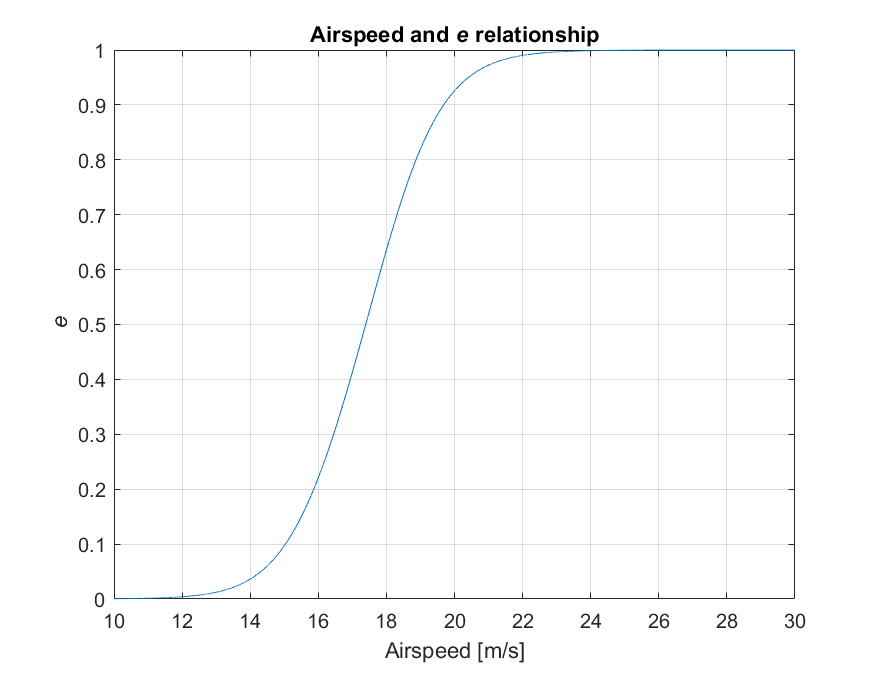}
        \caption{Effectiveness of aerodynamic with respect to Airspeed}
        \label{fig:my_label}
    \end{figure}

    \subsubsection{Aerodynamic Moments} 
    The aerodynamic moment denoted as $\mathbf{M}_a$ combines moments generated from the aileron, elevator, and rudder. Moments in roll, pitch yaw are each represented as $\mathbf{L}^\mathcal{B}$,$\mathbf{M}^\mathcal{B}$, and $\mathbf{N}^\mathcal{B}$. 
    \begin{equation}
        \mathbf{M}_{a} = \begin{bmatrix}L^\frm{B} \\ M^\frm{B} \\ N^\frm{B} \end{bmatrix}
    \end{equation}
    where

    \begin{align}
        \mathbf{L}^\mathcal{B} \approx \Bar{q}S_abC_L\delta_a, \quad
        \mathbf{M}^\mathcal{B} \approx \Bar{q}S_e\Bar{c}C_M\delta_a, \quad
        \mathbf{N}^\mathcal{B} \approx \Bar{q}S_rbC_N\delta_a
    \end{align}
    
In the equation, the wing surface area dedicated to the aileron, elevator, and rudder is denoted as $S_{a,e,r}$. The dynamic pressure is represented by $\Bar{q}$, while the mean aerodynamic chord is denoted as $\Bar{c}$. The wingspan is given by the symbol $b$. In the meanwhile, the effectiveness of the aileron, elevator, and rudder is represented by $C_{L}$, $C_{M}$, and $C_{N}$, respectively. These coefficients are typically determined through wind tunnel testing, providing information about the influence of the aileron, elevator, and rudder on the aerodynamics of the aircraft. In our simulation, these constants are: $S_a$ = 0.036m$^2$,  $S_e$ = 0.1364m$^2$,  $S_r$ = 0.004m$^2$, $b$ = 2m, $\Bar{c}$ = 0.22m, $C_{L}$ = 0.1173, $C_{M}$ = 0.556, and $C_{N}$ = 0.0881.

\subsubsection{Other forces and moments}

Other forces and moments exerted on our vehicle include the gravitational force $\Force{g} = \RBI\begin{bmatrix}0\\0\\mg\ZI\end{bmatrix}$ and the resisting moments $\Moment{g} = \sum_{i = 1}^4 \Torque{i}^\frm{B} =
    c_K\sum_{i = 1}^4 (-1)^{d_i} \omega_i^2 \begin{bmatrix}\sin {\chi_i}\\0\\-\cos{\chi_i}\end{bmatrix}$


The total forces exerted on the VTOL are determined by summing up all three types of forces: $\Force{} = \Force{r} + \Force{a} + \Force{g}$, where $\Force{r}$ is the thrust force, $\Force{a}$ is the aerodynamics force and $\Force{g}$ is the force generated by gravity. The total moments applied to the vehicle could be calculated as $\Moment{} = \Moment{r} + \Moment{g} + \Moment{a}$, Where $\Moment{r}$ is the thrust moment, $\Moment{a}$ is the aerodynamics moment and $\Moment{g}$ is the resisting moment.
    


\section{Control} \label{sec:control}
This section explains in detail the Model Predictive Control (MPC) controller and the control allocation strategy. The overall control architecture is shown in Fig.~\ref{fig:control_arch}.

\subsection{Control Architecture}
\begin{figure*}[h]
    \centering
    \includegraphics[width=\linewidth]{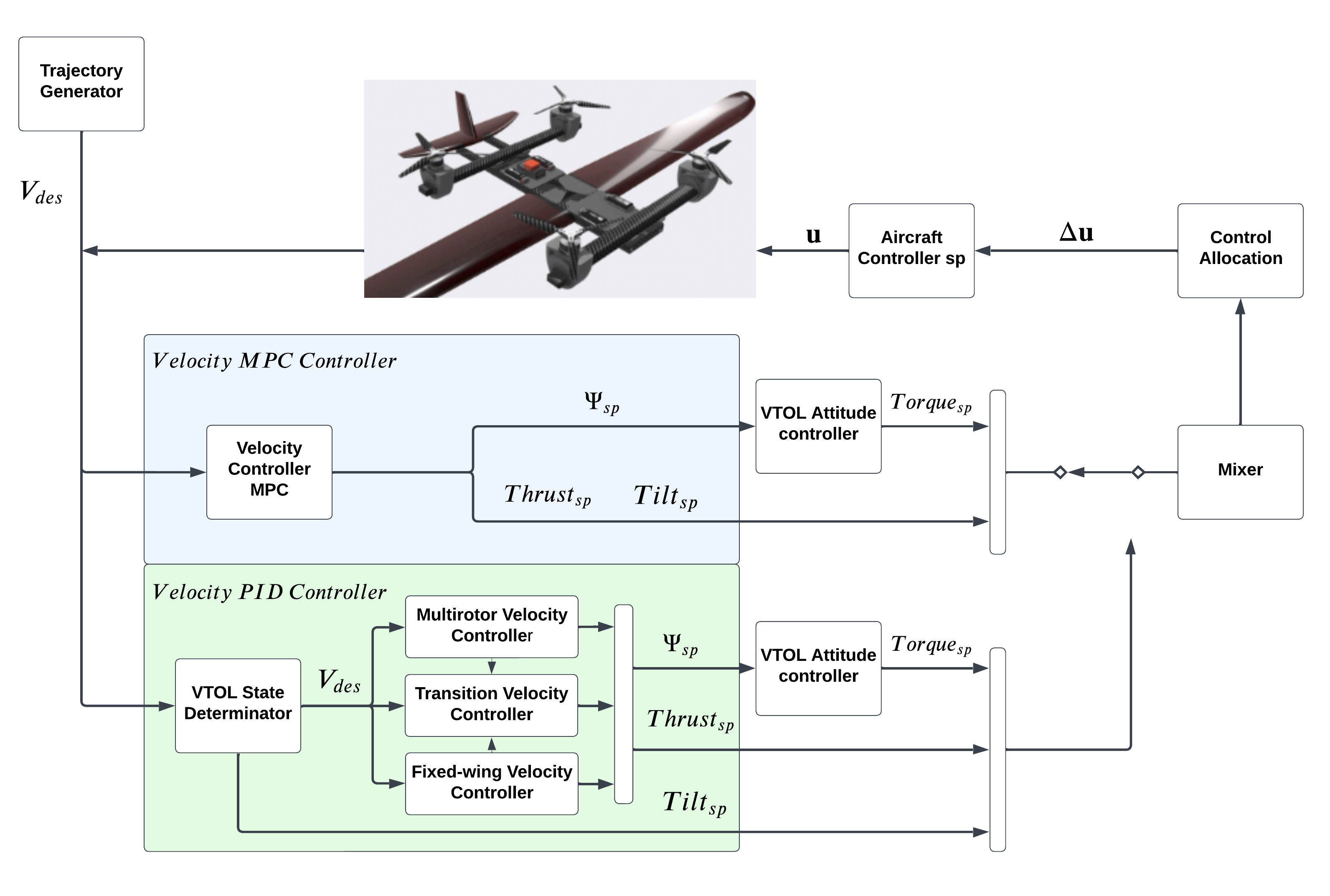}
    \caption{Overview of the proposed control architecture for the tiltrotor VTOL}
    \label{fig:control_arch}
\end{figure*}

The control architecture consists of the velocity controller, attitude controller, and control allocation module. (as shown in the Fig.~\ref{fig:control_arch}). The velocity controller takes in the velocity set points $V_{des}$ and calculates the desired attitudes $\Psi_{sp}$ to the VTOL attitude controller and outputs tilt angles and thrust set point $T_{sp}$. Control allocation maps these desired forces and torques into each of the individual actuation commands.  We also implement the approach as Pixhawk does using mode switching and PID control as a backup and comparison.

\subsubsection{Velocity Controller}
The outer velocity loop translates desired velocity set points into attitude set points, thrust set points, and tilt angles. The velocity MPC controller is designed to fully utilize the configuration redundancy. It offers a unified control strategy that does not require the identification of flight configurations. While solving the required attitude and thrust based on specific velocity set point inputs, the MPC controller also considers a variety of constraints which will be discussed later in the paper. For backup and comparison purposes, we also provide the approach Pixhawk does. It consists of a multirotor mode and a fixed-wing mode, either running separately in the corresponding VTOL phases or together with an attitude blender during the transition. 





\subsubsection{Attitude Controller}
After receiving the attitude setpoint from the outer loop, the attitude calculates a torque triplet so that the vehicle attitude can be maintained close to the setpoint. A PID controller is designed to determine the appropriate torque needed. The same VTOL attitude controller is utilized following the MPC or PID velocity controller. The torque setpoint together with previously calculated linear acceleration and tilt setpoint are then fed into the control allocation.

\subsubsection{Control Allocation}
The control allocation distributes the control wrench $\mathbf{W}^\mathcal{B}$ obtained from the outer loop and attitude controller to each actuator based on the aircraft's current state and configuration. This ensures that the overall force and torque acting on the vehicle are kept balanced and consistent. We implement a quadratic programming method to solve the allocation. The desired tilt angle is directly solved by the velocity MPC controller or VTOL state determinator.

\subsection{Model Predictive Control}
The MPC controller unifies the control irrespective of whether the vehicle is operating in multirotor or fixed-wing mode. The goal for MPC is to get an optimal sequence of U of inputs that minimizes the cost function of $J(x,u)$ over a horizon of $N$.

\begin{equation}
\begin{aligned}
\ U = \argmin_{u_k} \quad & \sum_{k=0}^{N}{J(x_k,u_k)}\\
\textrm{s.t.} \quad & x_{k+1} = f(x_k,u_k)\\
  \quad & x_k \in \chi    \\
  \quad & u_k \in \scalebox{1.4}{u}(x_k)   \\
\end{aligned}
\end{equation}

\subsubsection{Input and State Vectors}
The selection of input and state vectors is crucial considering the goal of unifying the fixed-wing and multi-rotor modes. The input should ideally be chosen to be physically feasible. Based on the system discussed in control architecture, input and state vectors for velocity MPC controllers are selected as follows:

\begin{equation}
\begin{aligned}
 \mathbf{x} = \begin{bmatrix} \mathbf{v} & \mathcal{X}&\Psi&\dot{\Psi}&\dot{\Psi}^- \end{bmatrix}^\top \in \mathbb{R}^{16\times1}\\ 
 \mathbf{u} = \begin{bmatrix}
 \dot{\mathcal{X}} & T & \Psi_{d}
 \end{bmatrix}\in \mathbb{R}^{8\times1}\\ 
\end{aligned}
\end{equation}

\noindent where the velocity $\mathbf{v}$ is in the inertial frame which makes it easier to track the global velocity trajectory generator. Attitude $\Psi$, attitude rate $\dot{\Psi}$, tilt angle $\mathcal{X}$ are chosen as states to optimize the controls: attitude $\Psi_{d}$, thrusts $T$ and tilt rates $\dot{\mathcal{X}}$. The MPC controller does not directly send torque setpoints to the control allocation; rather, it provides desired attitude setpoints $\Psi_{d}$ to the inner attitude control loop. Furthermore, the last attitude rate $\dot{\Psi}^-$ is to incorporate the attitude control loop into MPC controller for calculating the torque in the state update section. 

\par It is worth mentioning that in this system, the tilt angle and thrust of each rotor are treated independently. The tilt rate serves as a control input for the outer loop, while the tilt angles are considered as states. This approach allows for the integration of the physical constraints of the servo system. These constraints are associated with motor speeds and rotation angles of the servos, ensuring that the control inputs and states align with the practical limitations of the hardware.

\subsubsection{Constraints}
One advantage of the MPC controller is that it allows incorporating constraints that represent the physical limit of the output. The constraints applied are shown below:
\begin{equation}
\begin{aligned}
 \mathbf \lvert{\Psi}\rvert = \begin{bmatrix}\dfrac{\pi}{4},\dfrac{\pi}{4},\infty\end{bmatrix}^T rad, 
  \mathbf \lvert{\Psi_{d}}\rvert = \begin{bmatrix}\dfrac{\pi}{3},\dfrac{\pi}{3},\dfrac{\pi}{2}\end{bmatrix}^T rad\\
 \mathbf \lvert{\dot\Psi}\rvert = \begin{bmatrix}\pi,\pi,\pi\end{bmatrix}^T rad^{-1},
 \mathbf {\lvert{v}\rvert} = \begin{bmatrix}30,30,10\end{bmatrix}^T ms^{-1}\\
 T_{i} = \begin{bmatrix}0, 23\end{bmatrix}^T N,
 \mathbf{\mathcal{X}_{i}} = \begin{bmatrix}-7, 95\end{bmatrix}^T deg,
 \mathbf\lvert{\mathcal{\dot{X}}_{i}}\rvert = \begin{bmatrix}-\dfrac{\pi}{4}, \dfrac{\pi}{4}\end{bmatrix}^T rad^{-1}\\
\end{aligned}
\end{equation}

\subsubsection{Objective Function}
The design of the objective functions is the core component of the unified MPC strategy, which can automatically handle the flight phases to avoid mode switching. It takes many factors into account, such as trajectory tracking performance, state and input cost, and some soft constraints that have different sensitivity for low speed and high speed. In this way, the vehicle can automatically transit between different configurations rather than listening to explicit mode-switching commands. The objective function is designed as:
\begin{equation}
\begin{aligned}
J(\mathbf{x_k},\mathbf{u_k}) = J_{ref}(\mathbf {v}_{k},\mathbf{v}_{sp}) + J_{x,u}(\mathbf{x}_k,\mathbf{u}_k) + J_{soft}(\mathbf{x}_k)
\end{aligned}
\end{equation}

\begin{enumerate}
\item Reference tracking: The cost function of a typical squared error with ${Q}_{ref}$ is used to track the velocity set points. Additionally, when considering the performance of velocity tracking, particularly in the z-direction (e.g., preventing the vehicle from losing altitude or deviating from the trajectory), a final velocity cost with ${Q}_{f}$ is added.
\begin{equation}
\begin{aligned}
{J_{ref}(\mathbf {v}_{k},\mathbf{v}_{sp})} = {(\mathbf {v}_{k}-\mathbf{v}_{sp})}^T\mathbf{Q}_{ref}(\mathbf {v}_{k}-\mathbf{v}_{sp})+{(\mathbf {v}_N-\mathbf{v}_{sp})}^T\mathbf{Q}_{f}(\mathbf {v}_N-\mathbf{v}_{sp})
\end{aligned}
\end{equation}
\item State and control cost: $J_{x,u}(x_k,u_k)$ is the cost function using quadratic functions to calculate cost from states and controls based on the Table.~\ref{tab:Para}. The weights of the states have been adjusted according to how well the controller performed during the testing phase.
\begin{equation}
\begin{aligned}
J_{\mathbf{x,u}}(\mathbf{x}_k,\mathbf{u}_k) = \Psi^T\mathbf{Q}_{\mathbf{\psi}}\Psi+ \dot{\Psi}^T\mathbf{Q}_{\dot{\mathbf{\psi}}}\dot{\Psi}+\mathbf{u}_k^T\mathbf{R}_u\mathbf{u}_k
\end{aligned}
\end{equation}
\begin{table}[!htb]
    \centering
    \caption{Parameters of quadratic function for cost functions}
    \label{tab:Para}
    \begin{tabular}{ccc}
        \hline\hline
        \textbf{Name} & \textbf{Parameter} & \textbf{Value}    \\ \hline
        \textbf{$Q_{ref}$} &  $\begin{bmatrix}
            v_x, v_y, v_z
        \end{bmatrix}$&  $\begin{bmatrix}
            20,10,50
        \end{bmatrix}$    \\
        \textbf{$Q_{\psi}$} &  $\begin{bmatrix}
           \mathbf{\phi},\mathbf{\theta},\mathbf{\psi}
        \end{bmatrix}$&  $\begin{bmatrix}
            10,20,10
        \end{bmatrix}$    \\
        $Q_{\dot\psi}$ & $\begin{bmatrix}
          \dot{p},\dot{q},\dot{r}
        \end{bmatrix}$&  $\begin{bmatrix}
            3,3,3
        \end{bmatrix}$      \\
                $Q_{f}$ & $\begin{bmatrix}
          v_x, v_y, v_z
        \end{bmatrix}$&  $\begin{bmatrix}
            20,10,50
        \end{bmatrix}$    \\
        $R_u$ & $\begin{bmatrix}
          T,\dot{\mathcal{X}},\mathbf{\phi}_{sp},\mathbf{\theta}_{sp},\mathbf{\psi}_{sp}
        \end{bmatrix}$&  $\begin{bmatrix}
            0.025,1,10,20,10
        \end{bmatrix}$\\
    \hline
    \end{tabular}
\end{table}
\item Soft constraints: $J_{soft}(x_k)$ is to serve the purpose of constraining a reasonable tilt angle and velocity. It is mainly used to prevent special conditions of the vehicle tilting propellers forward at low speeds. However, at high speeds, all tilt angles are permitted.  Similar to the design in \cite{bauersfeld2021mpc}, an exponential function is designed as:
\begin{equation}
\begin{aligned}
J_{soft,\Bar{\mathcal{X}}}(v_{B},\Bar{\mathcal{X}}) = exp(a\cdot v_{B}\cdot \Bar{\mathcal{X}} + b\cdot \Bar{\mathcal{X}} + c\cdot v_{B} + d)\\
\end{aligned}
\end{equation}
The constants are changed to $ a = -0.33, b = 13.32, c = 1.8, d = -2.303$ for better performance of the solver. The values of the function are shown in the table \ref{tab:soft}. When the forward speed is low, the cost is significantly higher. This makes the propellers in a forward position difficult, making the servo angles away from 90 degrees. In addition, at the same low speed, if the aircraft behaves more similarly to the fixed-wing mode, having a higher servo angle, the cost will also be higher to discourage it in such a position. The main reason for these trends of costs is to save energy. When the airspeed is high, the vehicle will have a higher aerodynamic lift force. This reduces the need for the propeller to face upward to provide upward thrust. This encourages the aircraft to behave in fixed-wing which makes generating forward thrust for long-range tasks more efficient. 

\begin{table}[!htb]
    \centering
    \caption{Some values of the soft constraints cost function }
    \label{tab:soft}
    \begin{tabular}{ccccccc}
        \hline\hline
         \diagbox{\textbf{$\Bar{\mathcal{X}}$(deg)}}{\textbf{V(m/s)}} & \textbf{0} & \textbf{5} & \textbf{10} & \textbf{15}  \\ \hline
        \textbf{$0$} &  0.10 & 0.0092 & $8.40e^{-4}$ & $3.12e^{-4}$   \\
        \textbf{$45$} &  $1.50e^{34}$ & $20.1$ & $6.67e^{-33}$ & $2.22e^{-6}$ \\
        $90$ & $2.30e^{69}$ & $2.7e^{3}$ & $3.20e^{-23}$ & $0.00$ \\
        \hline
    \end{tabular}
\end{table}

\end{enumerate}

\subsubsection{State Update}
With known total force $\mathbf{F}^{\mathcal{I}}$ and torque $\boldsymbol{\tau}^{\mathcal{B}}$ acting on the vehicle, the state update can be easily obtained as described in Sec.\ref{sec:problem}. Notice that the inner-loop controller receives attitude setpoints from the MPC, as illustrated in Fig.~\ref{fig:control_arch}. Thus, a Proportional (P) controller is designed first to compute an attitude rate setpoint, taking into account both the current attitude and the pre-determined setpoint.
\begin{equation}
\begin{aligned}
 \dot{\mathbf{\Psi}}_{sp} = \mathbf{K}_p(\mathbf{\Psi}_{sp}-\mathbf{\Psi})  
\end{aligned}
\end{equation}
Then, a PD controller computes the desired torque.
\begin{equation}
\begin{aligned}
 \boldsymbol{\tau} = \mathbf{K}_p(\dot{\mathbf{\Psi}}_{sp}-\dot{\mathbf{\Psi}})+\mathbf{K}_d(\dot{\mathbf{\Psi}}^--\dot{\mathbf{\Psi}})
\end{aligned}
\end{equation}
The state update in continuous time can be written as:
\begin{equation}
 \dot{x}= \begin{bmatrix} \frac{1}{m}\cdot(\mathbf{T} +\mathbf{F}^{\mathcal{I}})+9.81\mathbf{e}_z \\ \dot{\mathcal{X}} \\ \dot{\Psi} \\ \mathbf{I}^{-1}\boldsymbol{\tau}^{\mathcal{B}} \end{bmatrix} \\ 
\end{equation}
where $\mathbf{F}^{\mathcal{I}}$ and $\boldsymbol{\tau}^{\mathcal{B}}$ are the total force and torque acting on the vehicle modeled from Sec.\ref{sec:problem}. $\mathbf{I}^{-1}$ is the inverse of the vehicle moment of inertia matrix. $\dot{\mathbf{\Psi}}^-$ is replicated from the previous iteration.

\subsection{Control Allocation}
Control allocation will distribute controller wrench output to the actuators. These setpoints include the rotation speed for each of the rotors as well as the aileron, elevator, and rudder deflection rate. In our approach, we utilize small perturbation theory to linearize the system dynamics around the trimmed state. A first-order perturbation analysis is employed. The actuator setpoint is represented by its difference from the actuator setpoint in the trim condition. 
\begin{equation}
\mathbf{u}_\mathrm{sp} = \mathbf{u}_\mathrm{trim} + \Delta\mathbf{u}
\end{equation}

To adapt VTOL's multiple flight mode, the trim state is chosen to be the actuator setpoint in the previous time step.
The previous velocity and attitude controller will generate a linear acceleration and an angular acceleration command. Since the aircraft weight and moment of inertia are known, these setpoints can be changed into a single wrench setpoint. To calculate the change in wrench setpoint from the current wrench, the current wrench can be calculated by 
\begin{align}
    \mathbf{W}^\mathcal{B} = \mathbf{A}\Delta\mathbf{u}+\mathbf{W}^\mathcal{B-}
\end{align}
where \textbf{A} is the control allocation's \textit{effectiveness matrix}, $\Delta$\textbf{u} is the change of actuator setpoint from the previous time step, and ${\mathbf{W^{\mathcal{B-}}}}$ is the wrench achieved by actuator setpoint in the previous time step. 

The effectiveness matrix is calculated by first acquiring the function of how actuators influence the generated wrench which is presented in the dynamics section. Taking the jacobian of the function will generate the matrix. In the effectiveness matrix, some variables such as rotor position, left aileron, right aileron, elevator, and rudder effectiveness coefficient are predefined by the hardware condition of our aircraft. The desired tilt solved by the MPC velocity controller will directly go into servo control and apply to the aircraft without going through control allocation. The current tilt determines the tilt angle in the effectiveness matrix. With the current generated wrench generated, its difference from the required wrench can be calculated.
\begin{align}
    \Delta\mathbf{W}^\mathcal{B} = \mathbf{W}_{\mathrm{sp}}^\mathcal{B} - \mathbf{W}^\mathcal{B}
\end{align}
where $\mathbf{W}^{\mathcal{B}}_{\mathrm{sp}}$ is the generated wrench by the actuators in the last time step. \\

For a tiltrotor VTOL UAV that possesses more actuators than necessary, the system becomes underdetermined, presenting an infinite array of solutions. This actuator redundancy also allows the aircraft to stay in its course while accommodating actuator failures. Two kinds of actuator failures are considered: rotor failure and control surface failure. To minimize the total effort exerted by the actuators, an optimization solver is formulated. It calculates the least-norm solution while satisfying the force and torque restrictions given by the velocity and attitude controller. A MATLAB optimization solver \textit{liqlin} is used to calculate the actuator setpoint difference. The optimization problem is represented as:
\begin{equation}\label{eq:optimization problem}
\begin{aligned}
    &\min_{\Delta\mathbf{u}}{J}(\mathbf{u}_\mathrm{sp}) \\
    &\textrm{s.t.} \quad {lb}\leq\Delta\mathbf{u}\leq{ub}
\end{aligned}
\end{equation}
$J$ is the objective function that we are trying to minimize:
\begin{equation}\label{eq:cost function}
    {J} = \frac{1}{2}||\mathbf{\mathbf{A}\Delta\mathbf{u}-\Delta\mathbf{W}^\mathcal{B}||_{2}^{2}}
\end{equation}
${lb}$ and ${ub}$ are the lower and upper bounds. Since the result of the solver is the actuator setpoint change, the bound is constantly changing based on the current actuator setpoint:
\begin{equation}
    {lb} = lb_{abs} - \mathbf{u}_\mathrm{trim}, \quad
    {ub} = ub_{abs} - \mathbf{u}_\mathrm{trim}
\end{equation}
$lb_{abs}$ and $lb_{abs}$ are the lower and upper bounds of the actuator setpoints. To improve the calculation speed of the solver, the bound for rotor speed is modified to 0 to 1, and the control surface is modified to -1 to 1. The lower and upper bounds are as follows:
\begin{align}
    lb_{abs} = \hspace{0.18em}
    \begin{bmatrix}
    0,& 0,& 0,& 0,& -1,& -1,& -1,& -1
    \end{bmatrix}, \quad
    ub_{abs} =
    \begin{bmatrix}
    1,& 1,& 1,& 1,& \hspace{0.65em}1,& \hspace{0.65em}1,& \hspace{0.65em}1,& \hspace{0.65em}1
    \end{bmatrix}
\end{align}

The first four elements of the matrix represent the rotor speed for each of the four rotors, while the last four elements denote the left aileron rate, right aileron rate, elevator rate, and rudder rate. In the event of complete actuator failures, both the lower and upper bounds for the actuators are set to zero. Additionally, the respective columns in the effectiveness matrix representing the broken actuators will also be set to zero to eliminate their influence in the calculation. If noncomplete failures occur and the actuators cannot be certain percentages more effective, the values in the bounds will reflect such failures. For example, if the second rotor cannot spin at more than 70\%, and the rudder cannot be more than 20\% effective, the modified bounds are:
\begin{align}
    lb_{abs} = \hspace{0.18em}
    \begin{bmatrix}
    0,& 0,& \hspace{0.75em}0,& 0,& 0,& -1,& -1,& -0.2
    \end{bmatrix}, \quad
    ub_{abs} =
    \begin{bmatrix}
    1,& 0.7,& 1,& 1,& 1,& \hspace{0.65em}1,& \hspace{0.65em}1,& \hspace{0.65em}0.2
    \end{bmatrix}
\end{align}

When a failure occurs on an actuator, other actuators will try to compensate for the loss in force and torque while maintaining the overall force and torque setpoints. However, the compensation is not always possible. When an actuator failure requires other actuators to over saturate, the allocation will not be able to accommodate such failure and will only try to achieve the desired force and torque as much as it can. 

The incorporation of limits in the MPC controller also addresses the problem of actuator saturation. The controller can avoid generating commands that exceed the actuator's operational range or violate predefined limits. For instance, it will not issue a command that causes the rotor to rotate beyond the maximum allowable speed.
\section{Simulation and Results} \label{sec:sim}
\subsection{Implementation Detail}
\begin{figure*}[h]
    \centering
    \includegraphics[width=0.8\linewidth]{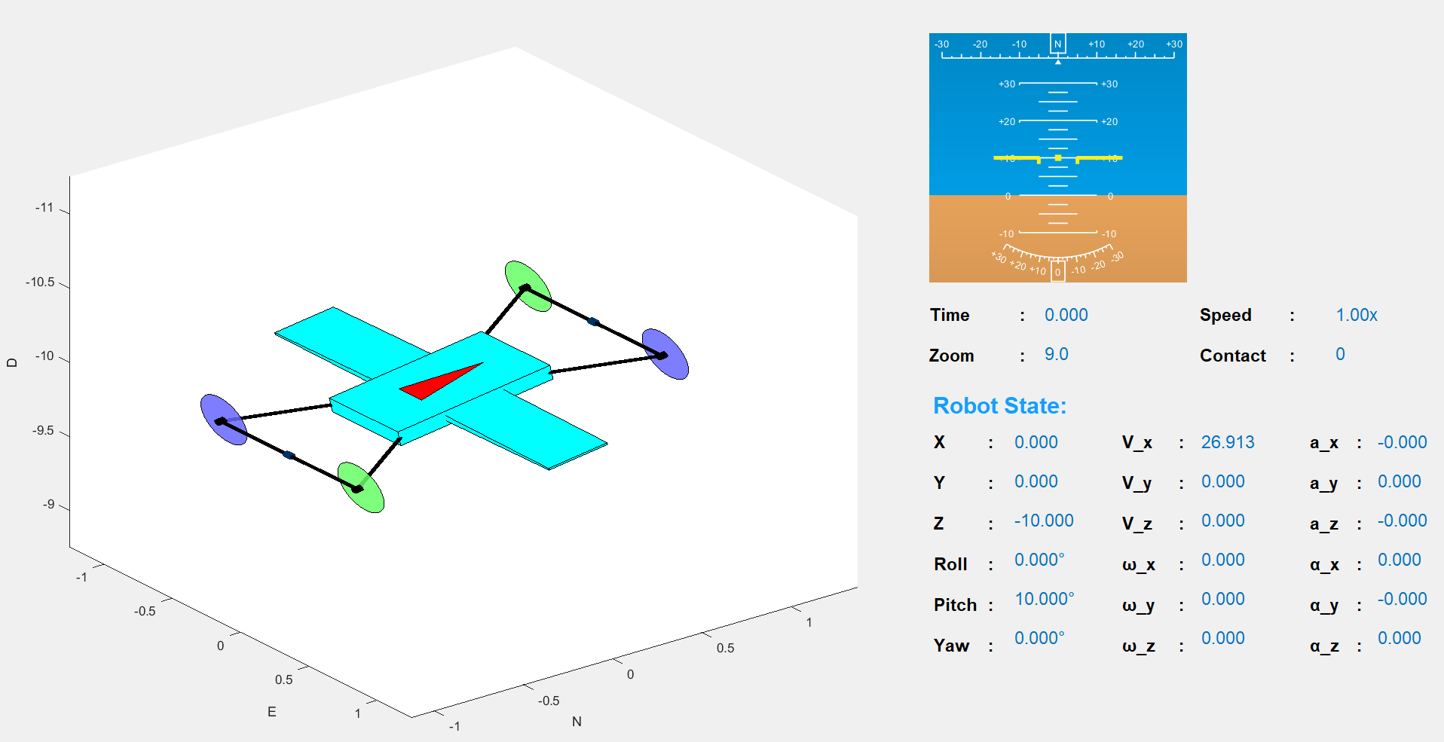}
    \caption{The GUI of our multi-purpose simulator.}
    \label{fig:sim}
\end{figure*}

We validate the proposed approach on a multi-purpose simulator developed by AirLab~\cite{Keipour:2022:unpub:simulator, Keipour:2022:thesis}. It uses object-oriented programming based on MATLAB language. Common UAV configurations can be created and simulated such as multirotor: quadrotor, hexarotor, octocopter, and VTOL. It can be used for various applications, e.g. aerial manipulation, design validation, wrench space analysis, etc, and thus can be used for rapid prototyping and control design for researchers. In this paper, we validate the developed control approach for the tiltrotor VTOL on this platform. Fig.~\ref{fig:sim} shows a visualization of the simulator GUI, where the aircraft is shown on the left and vehicle states and time are recorded on the right.

The simulation environment operates with a dynamic update frequency of 400Hz. The inner loop attitude controller functions at a frequency of 250Hz, while the outer loop MPC controller has a frequency of 100Hz. The prediction horizon N is set to 25, equivalent to 0.25s.

The MPC problem is formulated in nonlinear programs problem and solved by CasADi~\cite{Andersson2018}, which is an open-source tool for nonlinear optimization. Multiple shooting method is selected in the solver considering the solving speed as well as system convergence.  

\subsection{Results}
We present the results using the unified MPC control strategy for all flight phases including multirotor, fixed-wing phases, and the transition to validate the proposed control algorithm. We conducted a series of tests, starting with acceleration from hover to cruise speed while maintaining a fixed altitude. Subsequently, a scenario of acceleration and deceleration is designed, in which the aircraft accelerates from hover to cruise speed, followed by a transition back to hover. A test of acceleration from hover followed by moving in a circular path is designed to evaluate the system in handling turns and maneuvers along a uniform circular path. To evaluate the aircraft's ability to recover from actuator failures, two additional tests are conducted (1) the aircraft accelerates from the hover condition but one of the servos is jammed at a certain angle; (2) the aircraft simulates the process of a circular motion with one of the motors stops after the aircraft reaches a certain point. 

\subsubsection{Acceleration}
In this scenario, a fixed altitude acceleration trajectory is applied to the vehicle where velocity in $x$ direction accelerates from 0 m/s to 27.74 m/s and maintains at 27.74 m/s after. This range covers a wide range of velocities and includes all possible flight configurations, including both multirotor and fixed-wing modes. The trajectory given for both controllers can be described as:

\begin{equation}
\begin{aligned}
\mathbf{V}_{sp} =
  \begin{cases}

      \begin{bmatrix}2t,0,0\end{bmatrix} \ \ \text{if}\ V_{x,sp} < 27.74m/s \\
      \begin{bmatrix}27.74,0,0\end{bmatrix}  \ \ \text{if}\ V_{x,sp} \geq 27.74m/s
   \end{cases}
\end{aligned}
\end{equation}

\begin{figure} 
    \centering
    \subfigure[\label{fig:VX compare} Comparison of velocity tracking in x direction]{\includegraphics[width=0.49\linewidth]{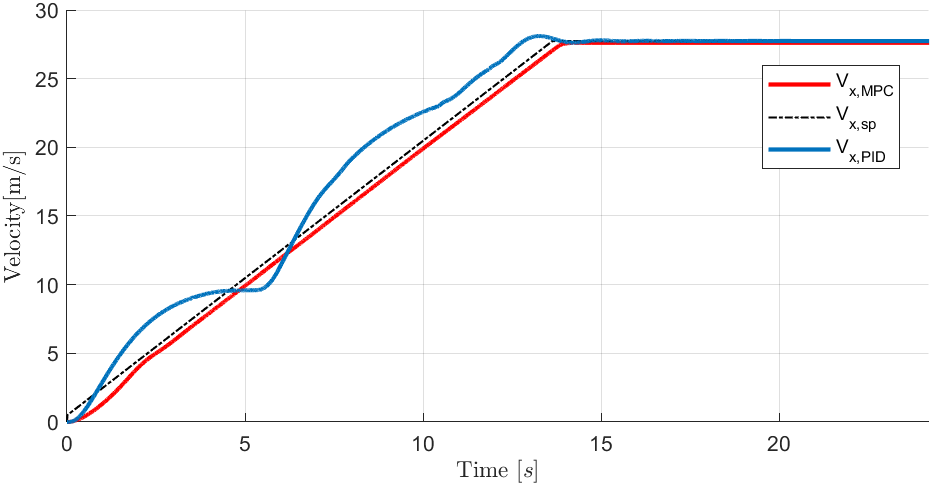}}
    \subfigure[\label{fig:VZ compare}  Comparison of velocity tracking in z direction]{\includegraphics[width=0.49\linewidth]{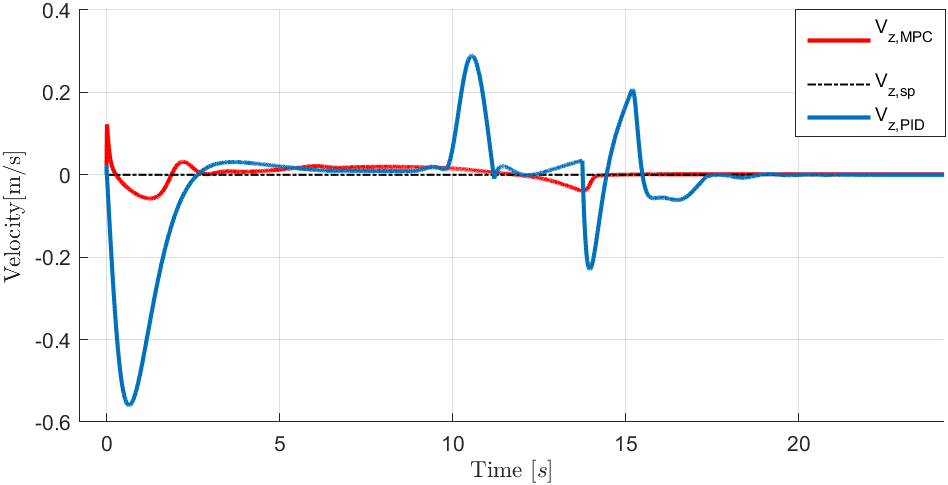}}
    \caption{ Comparison of velocity tracking in x and z direction of PID and MPC controller}
    \label{fig:velocityTack}
\end{figure}

Upon receiving a velocity trajectory, the airplane accelerates to the target velocity in approximately 14 seconds using both MPC and PID controllers. The result of trajectory tracking is shown in Fig.~\ref{fig:velocityTack}. The PID controller receives various velocity setpoints, including the multirotor max speed, which establishes the maximum speed for the VTOL in multirotor mode. The blending airspeed determines when the influence of the fixed-wing velocity controller becomes dominant, and the transition airspeed indicates when the VTOL is in fixed-wing mode. These setpoints guide the aircraft to follow a trajectory similar to the equation mentioned above. The MPC controller, on the other hand, is provided with an exact linear trajectory. During the entire simulation, the MPC controller does not fail to solve for the result. Thus, a switch back to the PID controller is not activated. In the acceleration phase, the airplane effectively maintains a near-zero vertical velocity. The MPC controller enables proactive adjustment of control signals, leading to smoother and more accurate velocity tracking. In addition, the MPC-based controller leads to less oscillating behavior with faster response and smaller overshoot. We also noticed that further tuning the gains of the PID controller cannot improve much of the control performance. During the transition phase, the PID control method lacks the ability to properly track velocity setpoint since the strategy is to conduct the transition as fast as possible. 

\begin{figure} 
    \centering
    \subfigure[\label{fig:tilt} Tilt angle]{\includegraphics[width=0.45\linewidth]{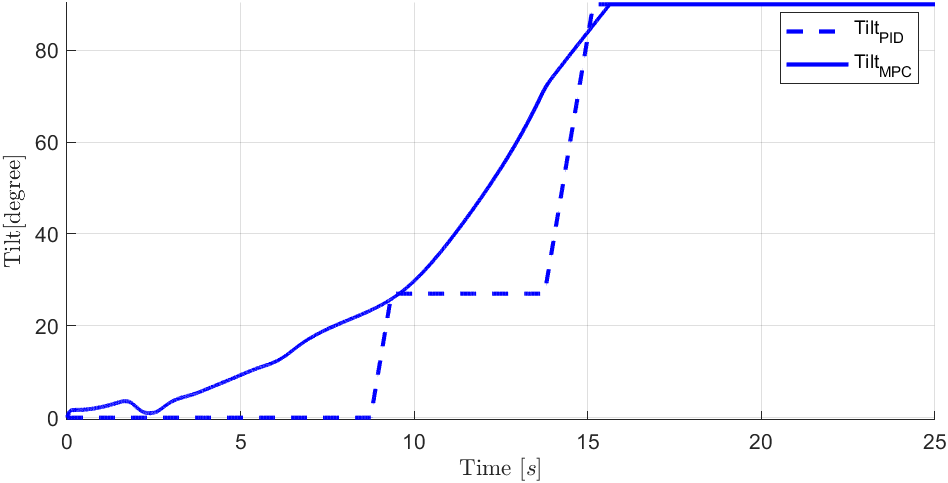}}
    \subfigure[\label{fig:pitch} Pitch angle]{\includegraphics[width=0.45\linewidth]{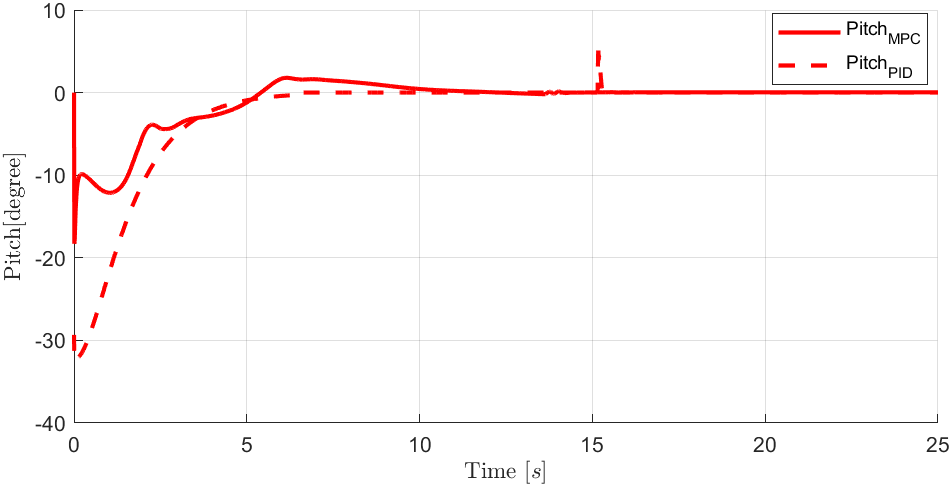}}
    \caption{Comparison of tilt angle and pitch angle of PID and MPC controller}
    \label{fig:tiltPitch}

\end{figure}

A comparison of the system's tilt and pitch response under PID and MPC controllers is presented in Fig.~\ref{fig:tiltPitch}. The MPC controller gradually increases the tilt from 0$^\circ$ to 90$^\circ$, resulting in a smooth and continuous transition. In contrast, the PID controller exhibits several segments of tilt caused by the VTOL mode switch during the transition. Additionally, the tilt change is more sudden and abrupt for the PID controller. As depicted in Fig.~\ref{fig:tilt}, the system demonstrates a smoother response when controlled by the MPC-based controller.

Similar trends are observed in the pitch response. Demonstrated in Fig.~\ref{fig:pitch}, the PID controller initiates a more aggressive approach by commanding a large pitch of -32$^\circ$ at the beginning, when the VTOL is gaining velocity in the $x$ direction. On the other hand, the MPC controller maintains a lower maximum pitch of -18$^\circ$, ensuring a smoother and more controlled change in pitch.

\subsubsection{Acceleration and Deceleration}
\begin{figure} 
    \centering
    \subfigure[\label{fig:MPCVelocity}Velocity tracking]{\includegraphics[width=0.45\linewidth]{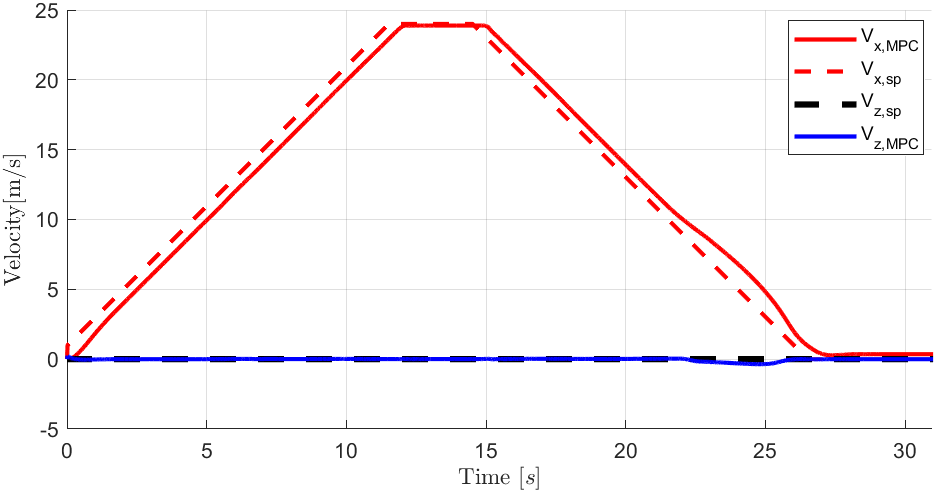}}
    \subfigure[\label{fig:MPCTilt}Pitch and tilt angle]{\includegraphics[width=0.45\linewidth]{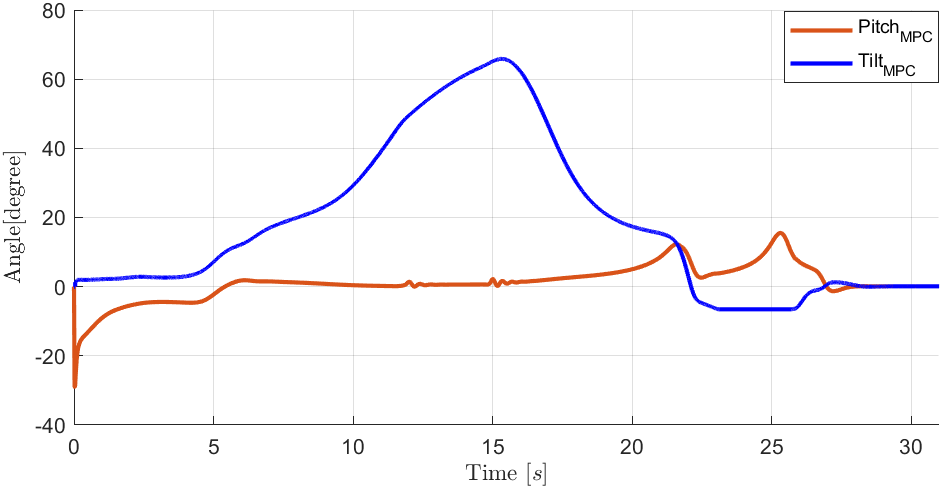}}
    \caption{MPC control performance during acceleration and deceleration}
    \label{fig:MPC_trans}
\end{figure}

Fig.~\ref{fig:MPC_trans} displays the outcome of an additional simulation in which VTOL accelerates from hovering and transits back to hovering. Like the previous acceleration simulation, set $V_{sp}$ at z direction to be zero during the whole process. The acceleration phase shows similar behavior as the previous test demonstrated in 
Fig.~\ref{fig:velocityTack} and Fig.~\ref{fig:tiltPitch}. Another observation is that the MPC controller can use the tilting mechanism to enhance the deceleration. Around 22 seconds, the VTOL tilts the motor backward at an angle of -7 degrees. Meanwhile, the aircraft is maintaining zero vertical speed. Once the vehicle comes to a stop, it tilts the propellers forward to balance the pitch of the aircraft.
This result also demonstrates the effectiveness of the control allocation method, showcasing its capability to distribute the desired torques and forces to the actuators in both acceleration and deceleration flight conditions.

\subsubsection{Uniform circular flight}
To test the VTOL's ability to perform turns in fix-wing mode, a flight task consisting of two parts is formulated. In the first part, the aircraft starts at a hover condition and gains airspeed at 2 m/s$^2$ until reaching 26 m/s. This segment is similar to the task presented in Fig.~\ref{fig:velocityTack}. After reaching 26 m/s, the second part begins with the aircraft performing a circular flight with a radius of 250 m with a constant velocity of 26 m/s until reaching a final time of 25 seconds. 

\begin{figure} [H]
\centering
\subfigure[\label{fig:Circletracking} Velocity tracking]{\includegraphics[width=0.45\linewidth]{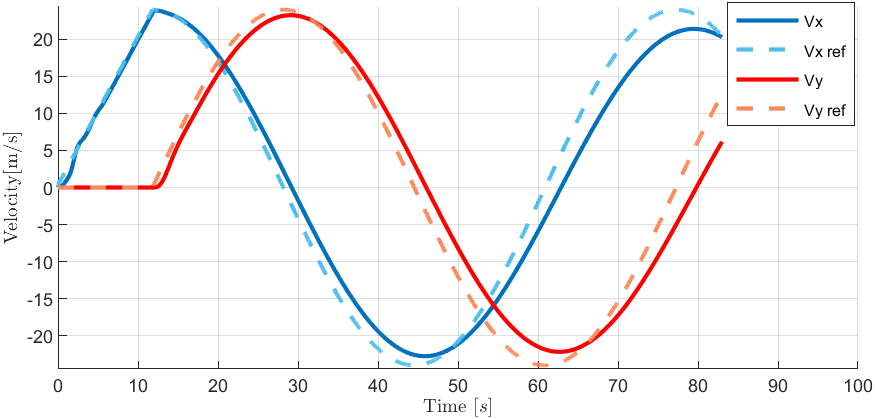}}
\subfigure[\label{fig:cicular_roll}Roll and Pitch angle]{\includegraphics[width=0.45\linewidth]{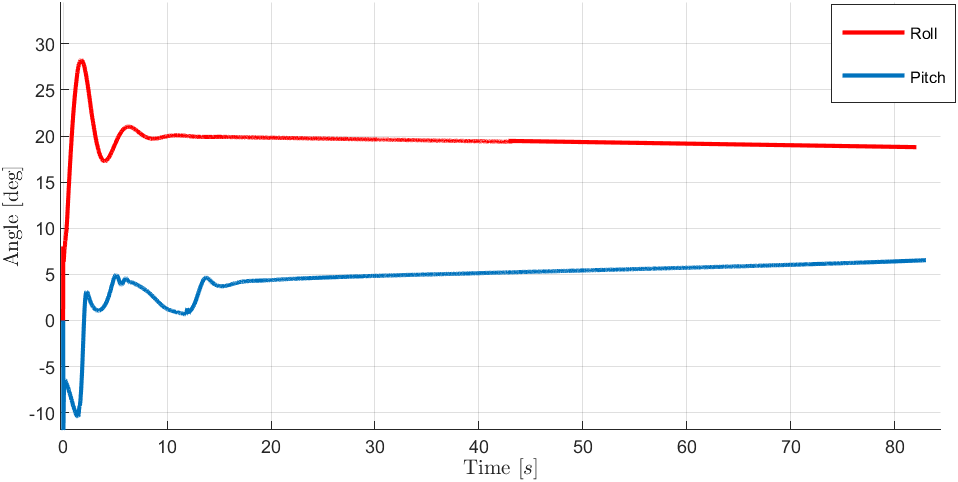}}
\caption{Velocity tracking performance and roll and pitch angle behaviors during uniform circular flight }\label{fig:uniformcicular}
\end{figure}

\begin{figure} [H]
\centering
\includegraphics[width=0.40\linewidth]{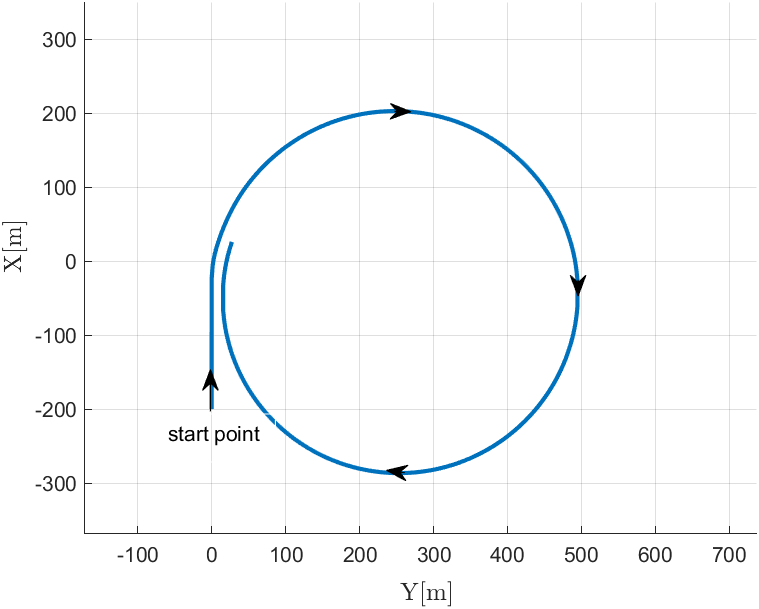}
    \caption{Trajectory of circular flight}
    \label{fig:cicularTraj}
\end{figure}

The velocity setpoints tracking performance and the aircraft roll and pitch behaviors are demonstrated in Fig.~\ref{fig:uniformcicular}. The overall position trajectory is shown in Fig.~\ref{fig:cicularTraj}. During the first part, the airspeed increases, and the aircraft presents very similar behavior to the flight task presented in Fig.~\ref{fig:velocityTack} and Fig.~\ref{fig:tiltPitch}. In the second part of the circular flight tasks, the MPC controller demonstrates its great performance in tracking the desired velocity. The aircraft holds a roll angle of 20 degrees to provide a centripetal force while pitching up to compensate for the loss in lift, and the yaw angle increases steadily to achieve the required velocity setpoints.   

\subsubsection{Actuators Failure}
To demonstrate the benefit of the system redundancy arising from the independent tilts, we examine two scenarios of actuator failures: (1) a total failure of one of the tilt servos during acceleration after hovering, and (2) a total failure of one of the motors during circular trajectory flight task. In scenario (1), during the procedure of gaining airspeed, the vehicle speed increases from 0 m/s to 27.74 m/s and maintains at this speed until reaching 25 seconds. This flight task is the same as the one described in Fig.~\ref{fig:VX compare}~\ref{fig:VZ compare} but with servo 1 jammed at 60 degrees, while other actuators continued to operate normally.

\begin{figure} [H]
    \centering
    \subfigure[\label{fig:tilt_set1}Servo 1 and 2 command]{\includegraphics[width=0.35\linewidth]{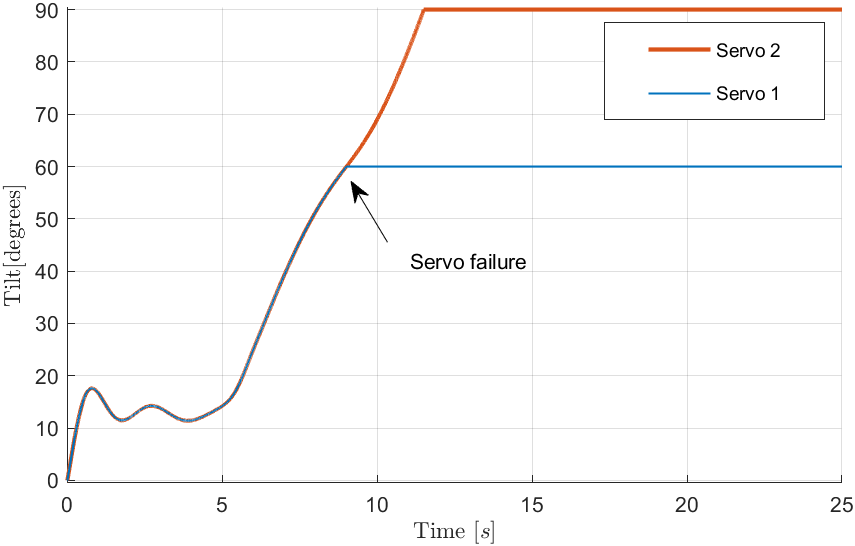}}
    \subfigure[\label{fig:tilt_set2}Servo 3 and 4 command]{\includegraphics[width=0.35\linewidth]{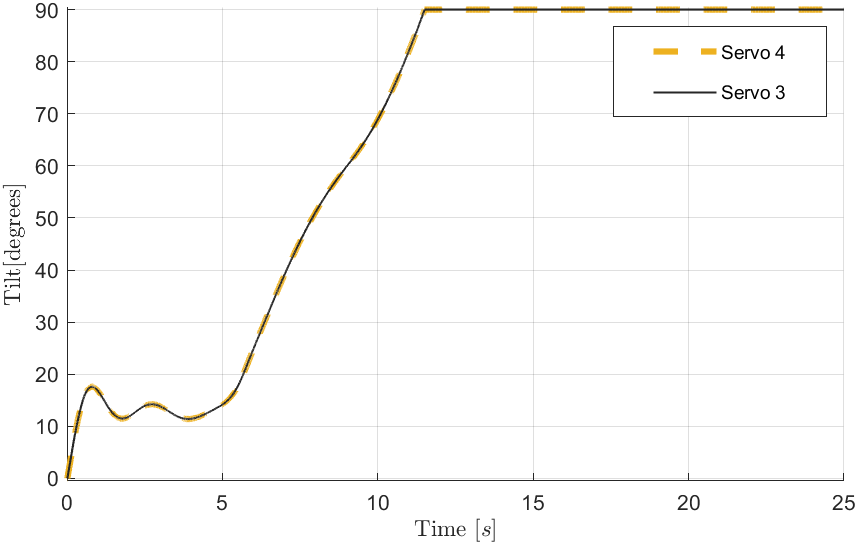}}
    
    \subfigure[\label{fig:tilt_motor12}Motor 1 and 2 command]
    {\includegraphics[width=0.35\linewidth]{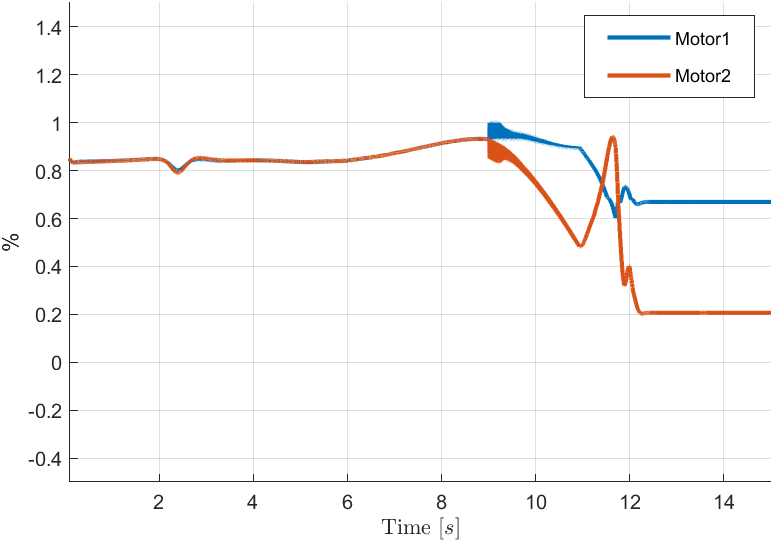}}
    \subfigure[\label{fig:tilt_motor34}Motor 3 and 4 command]
    {\includegraphics[width=0.35\linewidth]{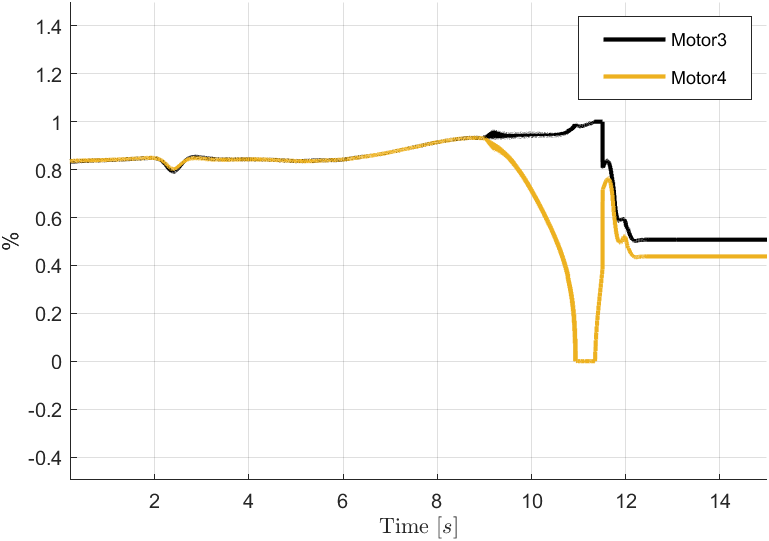}}
    \caption{Servo and motor command during a servo failure}
    \label{fig:tilt_fail}
\end{figure}

\begin{figure} [H]
    \centering
    \subfigure[\label{fig:tilt_vx}Velocity tracking in x direction]
    {\includegraphics[width=0.35\linewidth]{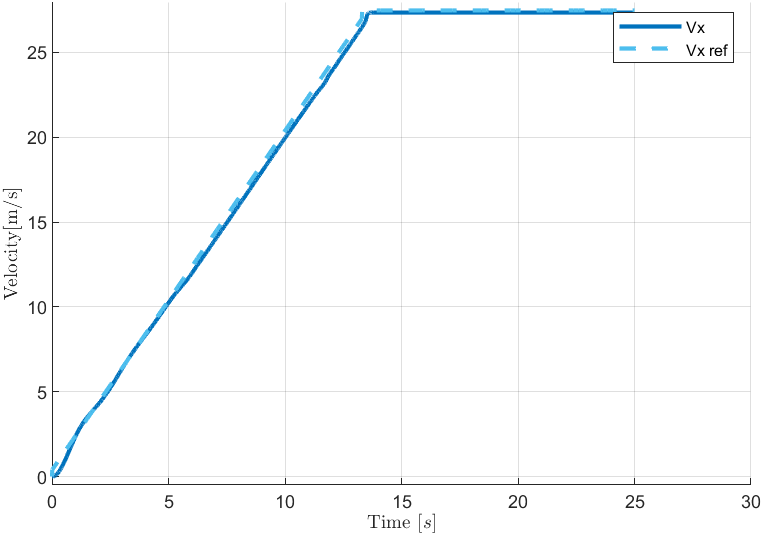}}
    \subfigure[\label{fig:tilt_vz}Velocity tracking in z direction]
    {\includegraphics[width=0.35\linewidth]{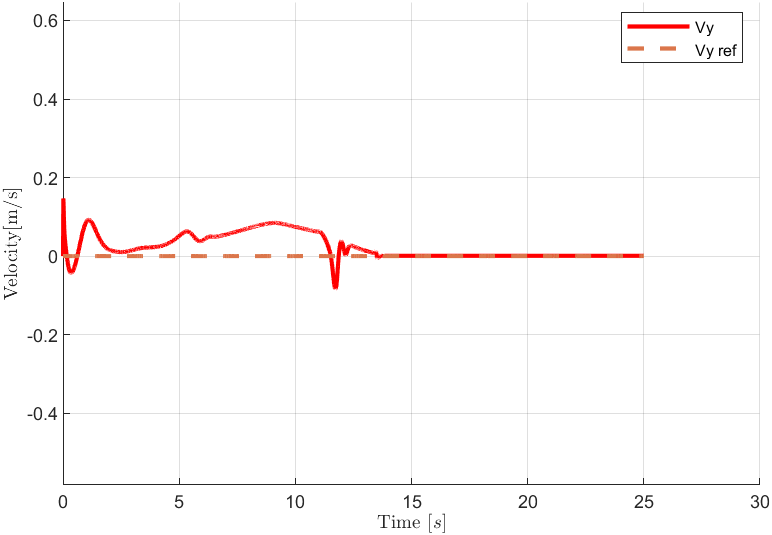}}
    
    \caption{Velocity tracking during a servo failure}
    \label{fig:tilt_failtracking}
\end{figure}

As shown in Fig.\ref{fig:tilt_fail}, the VTOL maintains a consistent pattern of increasing all four servo angles during acceleration. When the servo failure occurs, the remaining servos continue the same trend of increasing their angles. Even with a servo failure causing a brief deviation in roll and yaw angles, the MPC controller excels in tracking the desired velocity during circular flight tasks, restoring the aircraft's original course. In Fig.\ref{fig:tilt_failtracking}, the VTOL demonstrates outstanding velocity tracking, swiftly correcting any deviation induced by the servo failure.

In scenario (2), we simulate the abrupt malfunction of a motor within the context of a circular flight. This flight task is similar to the circular part of the uniform circular flight as described in Fig.~\ref{fig:uniformcicular} and Fig.~\ref{fig:cicularTraj}. The aircraft will maintain an airspeed of 26 m/s while flying in a circular path with a radius of 250 m. During the task, the first motor will cease to operate when the yaw angle reaches 90 degrees.

\begin{figure}[H]
\centering
\subfigure[\label{fig:motor_fail} Motor commands]{\includegraphics[width=0.45\linewidth]{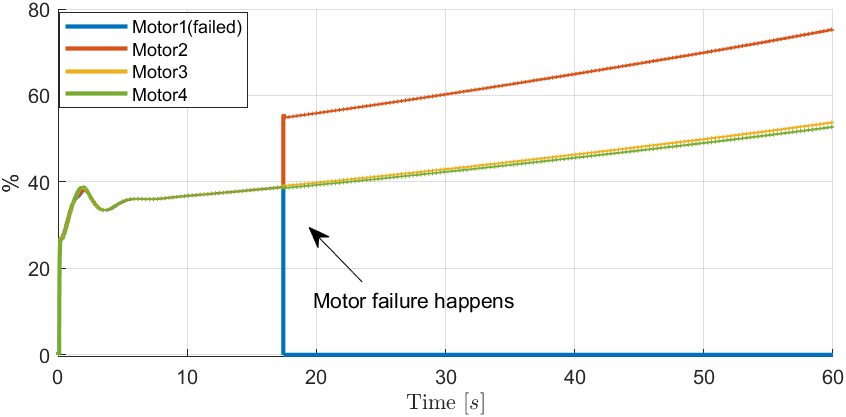}}
\subfigure[\label{fig:motor_servo} servo commands]{\includegraphics[width=0.45\linewidth]{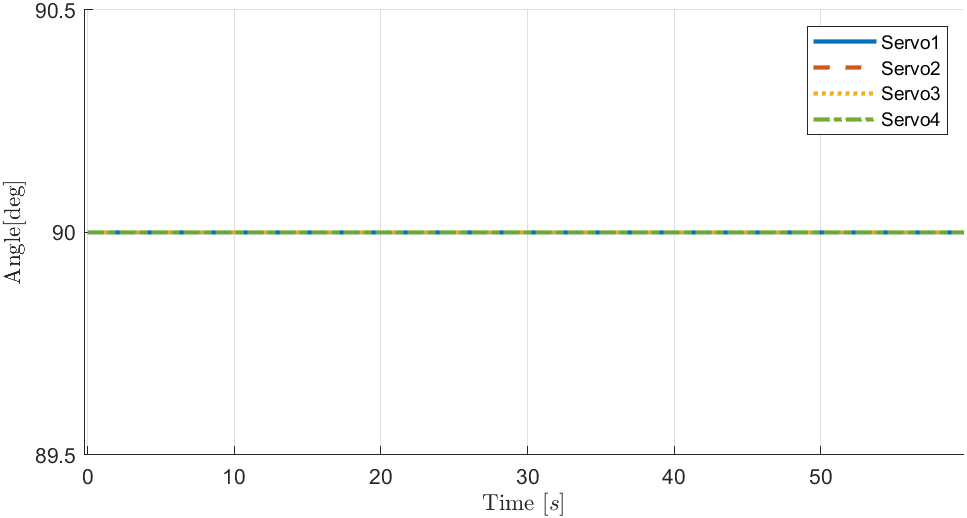}}

\caption{Motor commands and servo commands during motor failure}\label{fig:motor-failure}
\end{figure}

\begin{figure}[H]
\centering
\subfigure[\label{fig:motor_roll}Roll and roll setpoint]{\includegraphics[width=0.45\linewidth]{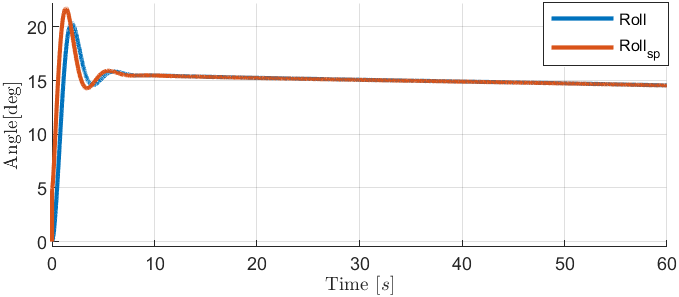}}
\subfigure[\label{fig:motor_yaw} Yaw and yaw setpoint]{\includegraphics[width=0.45\linewidth]{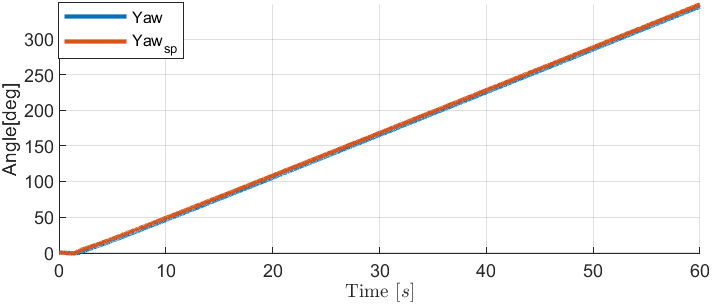}}

\caption{Roll and yaw angles during motor failure}
\label{fig:motor_rollyaw}
\end{figure}

\begin{figure}[H]
    \centering
    \includegraphics[width=0.38\linewidth]{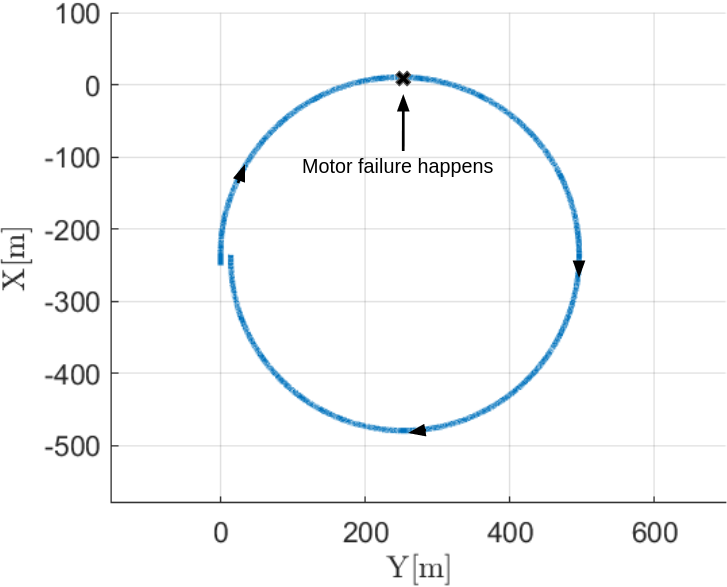}
    \caption{Trajectory of the VTOL of the motor failure test}
    \label{fig:motor_fail_2}
\end{figure}

The VTOL demonstrates similar behaviors as the uniform circular flight before the motor breaks down. As shown in Fig.~\ref{fig:motor-failure} and Fig.~\ref{fig:motor_fail_2}, motor 1 ceases to operate when the aircraft reaches 90 degrees yaw angle, dropping its spin rate to 0\%. This causes a sudden change in other actuator outputs as they are accommodating the system from such failure. All motors, especially motor 2, increase their speed after the motor failure. In Fig.~\ref{fig:motor_rollyaw}, aircraft shows very minimal errors in roll and yaw tracking. The system recovers quickly from motor failures.

In both scenarios, the results showcase the robustness and adaptability of our control system to mitigate the impact of actuator failure. Our system can consider actuator failures since it takes advantage of actuator redundancy. When an actuator ceases to operate, other actuators will offset the loss in force and torque. In addition, the control allocation running at a high frequency allows the system to respond faster when motor failures occur. 

While our VTOL can effectively handle certain types of failures, it is not capable of accommodating all potential failures. In the servo failure, the aircraft is gaining speed while maintaining its altitude. This procedure includes a transition from multirotor mode to fixed-wing mode. During the transition, if one servo is jammed at a lower angle, making it similar to its multirotor position, the aircraft will have difficulty in maintaining its flight task. In our observation, this is because jamming at a lower angle will reduce the aircraft's ability to generate thrust in the horizontal direction. This difficulty will lead to a maxing out of the broken servo's rotor speed which makes the tracking desired trajectory difficult.

In the motor failure test, in which the aircraft maintains a circular flight with a predefined radius, the value of the radius is essential in deciding if the aircraft can handle a single motor breakdown. During our tests, we observed that when the radius falls below 200m, our aircraft faces challenges in maintaining its course in the presence of a motor failure. This difficulty arises due to the increased need for centripetal acceleration caused by a lower turning radius, which demands a higher thrust. However, this increased thrust cannot be compensated by the aircraft with a malfunctioning motor, as the other actuators become saturated.

The key aspect of actuator failures is the exceeding control setpoints. They should not exceed the total capability of the remaining actuators. For example, if the controller is requesting a higher thrust than the remaining actuators' total thrust, the aircraft is unable to accommodate such failure. This is because the
dynamic update for the MPC controller will deviate too much from its predicted state and will lead to poor performance.

\section{Conclusion} \label{sec:conclu}
In this paper, a MPC-based control strategy for a specially designed tiltrotor VTOL is presented. This vehicle embodies a hybrid configuration, combining the characteristics of both fixed-wing aircraft and quadrotor aircraft. Prior work on controlling these aircraft either requires separate controllers and switching modes for different vehicle configurations or performs the control allocation on separate actuator sets, which cannot fully use the potential of the redundancy of tiltrotor. This paper introduces a unified MPC-based control strategy for a customized tiltrotor VTOL Unmanned Aerial Vehicle (UAV), which does not require mode-switching and can perform the control allocation in a consistent way. In addition, the redundancy in controls for the VTOL allows the controller to consider actuator failures and allows the system to recover from such failures. 

The proposed approach is implemented in a custom-developed multi-purpose simulator which can be used by researchers for rapid prototyping other vehicle designs or controller development. We present the simulation results for the various scenarios to validate the proposed MPC algorithm. Based on the results, the MPC demonstrates a smooth forward and backward transition by utilizing its tiltable propellers. It also shows a great ability to track circular trajectories. In addition, the system shows its distinctive ability to recover from both servo and motor failures during flight tasks, demonstrating the advantage of controlling the tilts independently.

\section*{Acknowledgments}
 We would like to express our sincere gratitude to Azarakhsh Keipour for his help with the multi-purpose simulator~\cite{Keipour:2022:unpub:simulator, Keipour:2022:thesis}
\bibliography{references}.

\end{document}